\documentclass[aos,MSNbibl,seceqn,citesort,dvips]{arximspdf}
\usepackage{dcolumn}
\usepackage{graphicx}

\doi{10.1214/11-AOS884}
\volume{39}
\issue{3}
\pubyear{2011}
\firstpage{1748}
\lastpage{1775}

\makeatletter
\newtheorem{proposition}{Proposition}[section]
\newproclaim{ass}{Assumption}
\newtheorem{theorem}{Theorem}[section]
\newcolumntype{d}[1]{D{.}{.}{#1}}
\makeatother

\begin{document}
\begin{frontmatter}

\title{A majorization--minimization approach to variable selection using spike and slab priors}
\runtitle{MAP estimation and variable selection}

\begin{aug}
\author{\fnms{Tso-Jung} \snm{Yen}\corref{}\ead[label=e1]{tjyen@stat.sinica.edu.tw}\thanksref{T1}}
\thankstext{T1}{Supported in part by Grants NSC 97-3112-B-001-020 and
NSC 98-3112-B-001-027 in the National Research Program for Genomic Medicine.}
\runauthor{T.-J. Yen}
\affiliation{Academia Sinica}
\address{Institute of Statistical Science\\
Academia Sinica\\
128 Academia Road, Section 2\\
Taipei 115\\
Taiwan\\
\printead{e1}} %adresu isvedimo komanda gale!
\end{aug}

% HISTORY:
\received{\smonth{7} \syear{2010}}
\revised{\smonth{12} \syear{2010}}

% ABSTRACT
%
\begin{abstract}
We develop a method to carry out MAP estimation for a class of Bayesian
regression models in which coefficients are assigned with
Gaussian-based spike and slab priors. The objective function in the
corresponding optimization problem has a Lagrangian form in that
regression coefficients are regularized by a mixture of squared $l_{2}$
and $l_{0}$ norms. A tight approximation to the $l_{0}$ norm using
majorization--minimization techniques is derived, and a coordinate
descent algorithm in conjunction with a soft-thresholding scheme is
used in searching for the optimizer of the approximate objective.
Simulation studies show that the proposed method can lead to more
accurate variable selection than other benchmark methods. Theoretical
results show that under regular conditions, sign consistency can be
established, even when the Irrepresentable Condition is violated.
Results on posterior model consistency and estimation consistency, and
an extension to parameter estimation in the generalized linear models
are provided.

\end{abstract}
%
% KEYWORDS
%

\end{frontmatter}
%
%s1 ###
\section{Introduction}\label{sec1}
Consider the following regression model:
%
%e1 ###
\begin{equation}
\label{1}
Y_{i} = x_{i1}\beta_{1}+x_{i2}\beta_{2}+\cdots+x_{ip}\beta_{p}+\varepsilon_{i},
\end{equation}
where $Y_{i}$ is the response variable for the $i$th subject, $x_{ij}$
is the $j$th covariate for the~$i$th subject, $\beta_{j}$ is the
corresponding regression coefficient and $\varepsilon_{i}$ is the error
term following some specified distribution. Variable selection in
regression problems has long been considered as one of the most
important issues in modern statistics. It involves choosing an
appropriate subset $\widehat{S}$ of indices $\{1,2,\ldots,p\}$ so that
for $j\in\widehat{S}$, the covariates $x_{ij}$'s and estimated
coefficients $\widehat{\beta}_{j}$'s are scientifically\vspace*{2pt} meaningful in
interpretation, and estimates $\widehat{y}_{i^{\prime}}=\sum_{j\in
\widehat{S}}x_{i^{\prime}j}\widehat{\beta}_{j}$ have relative good
properties in prediction.

In this paper, we develop a method to carrying out maximum a posteriori
(MAP) estimation for a class of Bayesian models in tackling variable
selection problems. The use of MAP estimation in variable selection
problems had previously been studied by Genkin et al.~\cite{genkinetal07} on logistic regression models with Laplace priors. The
difference between our model and Genkin et al.'s is that our model
assigns a Gaussian-based spike and slab prior weighted by Bernoulli
variables on each regression coefficient. Traditionally, parameter
estimation for this model and other Bayesian variable selection
settings relies on Markov chain Monte Carlo for posterior simulation
\cite{georgeandmcculloch93,georgeandmcculloch97,parkandcasella08,hans09,liandlin10,griffinandbrown10}
and empirical Bayes methods~\cite{georgeandfoster00,clydeandgeorge00,johnstoneandsilverman05}. A major
advantage of MCMC-based inference procedures is that they provide a
practical way to assessing posterior probabilities, and inference tasks
such as point estimation can be carried out straightforwardly based on
posterior probability calculation. However, convergence of MCMC-based
sampling algorithms is not often guaranteed and they may become
time-consuming as the number of covariates $p$ becomes quite large. A
different inference procedure on models with spike and slab priors is
recently provided by Ishwaran and Rao~\cite{ishwaranandrao05a,ishwaranandrao05b}, in that regression coefficients
are estimated via OLS-based shrinkage methods.

Our estimation method is different from the above approaches in several
aspects. In our MAP estimation, an augmented version of the posterior
joint density is derived. From frequentists' point of view, the MAP
estimation is equivalent to the regularization estimation with a
mixture penalty of squared $l_{2}$ and $l_{0}$ norms on regression
coefficients. In practice, we apply a majorization--minimization
technique to modify the penalty function so that convexity of the
objective function can be achieved. We then construct a coordinate
descent algorithm based on a specified iteration scheme to obtain the
MAP estimate. The algorithm involves iteratively applying
shrinkage-thresholding steps to obtain estimates that have sparse
features, that is, some of them have exact zero values. In this sense,
parameter estimation and variable selection can be achieved
simultaneously. In addition, the algorithm can be implemented
practically in a situation in which the number of covariates $p$ is
much larger than the number of samples $n$. It is different from the
OLS-based methods in that $p\leq n$ is required to avoid singularity in
matrix operation. The algorithm can also be fast when $p$ is large but
the number of covariates with nonzero coefficients is small. Simulation
studies show that the MAP estimate can lead to better performances in
variable selection than those based on other benchmark methods in
various circumstances.

Recent frequentists' approaches to variable selection focus on applying
the idea of regularization estimation in the situation in which the
number of variables is much larger than the number of samples
\cite{fanandli01,zouandli08,zouandhastie05,zouandzhang09,friedmanetal09,zou06,yuanandlin06,meieretal08,candesandtao07,meinshausen07,zhang10}.
All these approaches can either been seen as alternatives or as
extensions of the lasso estimation~\cite{tibshirani96}. For theoretical
properties of the lasso estimation, Knight and Fu~\cite{knightandfu00}
pointed out that with regular conditions on the order magnitude of the
tuning parameter, the lasso is consistent in parameter estimation.
However, as shown by Meinshausen and B\"{u}hlmann~\cite{meinshausenandbuhlmann06} and Zou~\cite{zou06}, for the lasso
estimation, consistency in parameter estimation does not imply
consistency in variable selection. Further conditions on the design
matrix and tuning parameter should be imposed to ensure consistency in
variable selection for the lasso estimation. In this aspect, Zhao and
Yu~\cite{zhaoandyu06} established the Irrepresentable Condition and
showed that the lasso can be asymptotically consistent in both variable
selection and parameter estimation if the Irrepresentable Condition
holds and some regular conditions on the tuning parameter are
satisfied. The same condition was also established by Zou~\cite{zou06}
and Yuan and Lin~\cite{yuanandlin07}. Later we will show that the MAP
estimator proposed in this paper is asymptotically consistent in
variable selection even when frequentists' Irrepresentable Condition is
violated.

The paper is organized as follows. Section~\ref{sec3} focuses
methodological aspects of the proposed method. Section~\ref{sec4}
provides two simulation studies on performances of the proposed method.
Section~\ref{sec5} develops relevant asymptotic analysis for the
method. Section~\ref{sec6} extends the method to parameter estimation
in the generalized linear models. Real data examples are provided in
Section~\ref{sec7}. Some concluding remarks are given in Section~\ref{sec8}.

%s2 ###
\section{Notation}\label{sec2}
Let $X$ be an $n\times p$ design matrix. Let $x_{i}$ denote the $i$th
row of~$X$ and $x_{ij}$ denote the $ij$th entry of $X$. The transpose
of $X$ is denoted by $X^{T}$. Let $y=(y_{1},y_{2},\ldots,y_{n})$ denote
the realization of random vector $Y=(Y_{1},Y_{2},\ldots,Y_{n})$ and
$\beta=(\beta_{1},\beta_{2},\ldots,\beta_{p})$ denote the regression
coefficient vector. Let $I_{p\times p}$ denote the $p\times p$ identity
matrix. For a $p$-dimensional vector $a=(a_{1},a_{2},\ldots,a_{p})$,
define the $l_{1}$ norm by $\Vert a\Vert _{1}=\sum_{j=1}^{p}|a_{j}|$, the
$l_{2}$ norm by $\Vert a\Vert _{2}=(\sum_{j=1}^{p}|a_{j}|^{2})^{1/2}$, the
$l_{\infty}$ norm by $\Vert a\Vert _{\infty}=\max_{j}|a_{j}|$ and the $l_{0}$
norm by $\Vert a_{j}\Vert _{0}=\sum_{j=1}^{p}\mathbb{I}(a_{j}\neq0)$, where
$\mathbb{I}(a_{j}\neq0)$ is an index variable such that $\mathbb
{I}(a_{j}\neq0)=1$ if $a_{j}\neq0$ and $\mathbb{I}(a_{j}\neq0)=0$
otherwise. The probability density of a random variable $Z$ conditional
on $\theta$ is denoted by $f(z| \theta)$. We define $S=\{j\dvtx  \beta
_{j}\neq0, j=1,2,\ldots,p\}$, that is, the index set of nonzero valued
coefficients in $\beta=(\beta_{1},\beta_{2},\ldots,\beta_{p})$. We
further define $X_{S}$ as the design matrix of $X$ whose columns are
indexed by $S$. Finally, we define the sign function for variable $z$
as $\operatorname{sign}(z)=1$ if $z> 0$; $\operatorname{sign}(z)=-1$ if $z<0$; $\operatorname
{sign}(z)=0$ if $z=0$.

%s3 ###
\section{The method}\label{sec3}
We start by assigning prior distributions on parameters in the
regression model (\ref{1}). Note that in a regression model a covariate
can only be selected if its coefficient is estimated with a nonzero
value. Based on this observation, we assign an index variable $\gamma
_{j}$ to each covariate and define that $\gamma_{j}=1$ if $\beta
_{j}\neq0$ and $\gamma_{j}=0$ if $\beta_{j}=0$. Here, we may write
$\gamma_{j}=\mathbb{I}(\beta_{j}\neq0)$. With the definition of $\gamma
_{j}$, the regression model (\ref{1}) has an equivalent representation
given by
\[
Y_{i} = \sum_{j=1}^{p} x_{ij}\gamma_{j}\beta_{j} + \varepsilon
_{i}.
\]
From a variable selection point of view, the index vector $\gamma
=(\gamma_{1},\gamma_{2},\ldots,\gamma_{p})$ is an indicator for
candidate models. Different candidate models will have different values
in $\gamma$.
%s3.1 ###
\subsection{The Bayesian formulation}
Under a Bayesian framework, we assume
%
%e2 ###
\begin{eqnarray}
\label{2}
Y_{i}| x_{i},\beta,\gamma,\sigma^{2} &\sim&\operatorname{Normal}\Biggl(\sum
_{j=1}^{p}x_{ij}\gamma_{j}\beta_{j},\sigma^{2}\Biggr)  \qquad \mbox{for
}i=1,2,\ldots,n,\nonumber\\
\beta_{j}| \sigma^{2},\gamma_{j},\lambda&\sim&\gamma_{j}\operatorname{Normal}(0,\sigma^{2}\lambda^{-1}) + (1-\gamma_{j})\mathbb{I}(\beta
_{j}=0)\nonumber\\
&& \hspace*{168pt}\mbox{for }j=1,2,\ldots,p,\hspace*{-168pt}
\\
\sigma^{2}| \tau_{1},\tau_{2}&\sim&\mathrm{Inverse\mbox{-}Gamma}(\tau_{1},\tau
_{2}),\nonumber\\
\gamma_{j}| \kappa&\sim&\operatorname{Bernoulli}(\kappa)  \qquad  \mbox{for
}j=1,2,\ldots,p.\nonumber
\end{eqnarray}
The prior distribution of $\beta_{j}$ given in (\ref{2}) is the spike
and slab prior originally proposed by Mitchell and Beauchamp~\cite{mitchellandbeauchamp88}. It implies that conditional on $\gamma
_{j}=0$, $\beta_{j}$ is equal to 0 with probability one, and
conditional on $\gamma_{j}=1$, $\beta_{j}$ follows a normal
distribution with mean $0$ and variance $\sigma^{2}\lambda^{-1}$. The
Bernoulli prior on $\gamma_{j}$ says that if only prior information is
available, $\gamma_{j}$ will have probability $\kappa$ to be 1 and
$1-\kappa$ to be 0. Note that since $\gamma_{j}\in\{0,1\}$, we can
express\vspace*{1pt} the mixture form of the prior on $\beta_{j}$ as $\operatorname{Normal}(0,\sigma
^{2}\lambda^{-1})^{\gamma_{j}}\times\mathbb{I}(\beta_{j}=0)^{1-\gamma
_{j}}$. This representation will be used in deriving the joint
posterior density of the parameters.

Under Bayesian model (\ref{2}), the joint posterior density of $\beta$,
$\gamma$ and $\sigma^{2}$ can be expressed as
%
%e3 ###
\begin{eqnarray}
\label{3}
& &f(\beta,\gamma,\sigma^{2}| X,y,\lambda,\tau_{1},\tau_{2},\kappa
)
\nonumber
\\[-8pt]
\\[-8pt]
\nonumber
& &\qquad        \propto f(y| X,\beta,\gamma,\sigma^{2})f(\beta| \sigma^{2},\gamma,\lambda)f(\sigma^{2}| \tau_{1},\tau_{2})f(\gamma| \kappa).
\end{eqnarray}
With (\ref{3}), we can estimate $(\beta,\gamma,\sigma^{2})$ via various
inference methods. In this paper, the \textit{maximum a posteriori} (MAP)
method is adopted. Formally, the MAP estimator for $(\beta,\gamma,\sigma
^{2})$ is defined by
%
%e4 ###
\begin{equation}
\label{5}
(\widehat{\beta},\widehat{\gamma},\widehat{\sigma}^{2})=\arg\min_{\beta
,\gamma,\sigma^{2}}\{-2\log f(\beta,\gamma,\sigma^{2}| X,y,\lambda,\tau
_{1},\tau_{2},\kappa)\},
\end{equation}
that is, the minimizer of the minus 2 logarithm of the joint posterior
density. The minus 2 logarithm of the joint posterior density\vadjust{\goodbreak} can be
explicitly expressed as
%
%e5 ###
\begin{eqnarray}
\label{method1}
-2\log f(\beta,\gamma,\sigma^{2}| X,y,\lambda,\tau_{1},\tau
_{2},\kappa)    &=&\frac
{1}{\sigma^{2}}\sum_{i=1}^{n}\Biggl(y_{i}-\sum_{j=1}^{p}x_{ij}\gamma
_{j}\beta_{j}\Biggr)^{2}\nonumber\\
&&{}  +\frac{\lambda}{\sigma^{2}}\sum_{j=1}^{p}\gamma_{j}\beta
_{j}^{2}
\nonumber
\\[-8pt]
\\[-8pt]
\nonumber
& & {}                           +\frac{2\tau_{2}}{\sigma^{2}}+(n + 2\tau_{1}+2)\log\sigma^{2}\nonumber\\
& &{}                            +\sum_{j=1}^{p}\gamma_{j}\log\biggl\{\frac{2\pi\sigma^{2}(1-\kappa
)^{2}}{\lambda\kappa^{2}}\biggr\} + \mathrm{const}.\nonumber
\end{eqnarray}
Here, we have used an equivalent representation $\operatorname{Normal}(0,\sigma
^{2}\lambda^{-1})^{\gamma_{j}}\times\mathbb{I}(\beta_{j}=0)^{1-\gamma
_{j}}$ for $f(\beta_{j}| \sigma^{2},\lambda,\gamma)$ given that $\gamma
_{j}\in\{0,1\}$. Note that in (\ref{method1}) the term $\sum
_{j=1}^{p}\log\mathbb{I}(\beta_{j}=0)^{1-\gamma_{j}}$
vanishes since for every $j$, $\gamma_{j}=1$ implies $\mathbb{I}(\beta
_{j}=0)=0$. In turn, $(1-\gamma_{j})\log\mathbb{I}(\beta_{j}=0)=0\cdot
\infty=0$. On the other hand, $\gamma_{j}=0$ implies $\mathbb{I}(\beta
_{j}=0)=1$, and in turn $\log\mathbb{I}(\beta_{j}=0)=\log1 = 0$.

For practical purposes, we fix $\sigma^{2}$ and multiply (\ref{method1}) with $\sigma^{2}$ in the following discussion. Given that
$\sigma^{2}$ is fixed, the function (\ref{method1}) has some
meaningful interpretations in terms of regularization estimation on
$\beta$. For example, by definition $\gamma_{j}\geq0$, and the
quantity $\sum_{j=1}^{p}\gamma_{j}$ can be seen as an $l_{1}$ norm on
the vector $\gamma$. Given the above argument, we can write the fourth
term on the right-hand side of (\ref{method1}) as $\rho_{\lambda,\kappa
,\sigma^{2}}\Vert \gamma\Vert _{1}$, where $\rho_{\lambda,\kappa,\sigma
^{2}}=\sigma^{2}\log[2\pi\sigma^{2}\lambda^{-1}(1-\kappa)^{2}\kappa
^{-2}]$. Note that as $\kappa$ increases, $\rho_{\lambda,\kappa,\sigma
^{2}}$ will decrease. It implies that a strong belief in the presence
of a variable will decrease the penalty value for the variable. In
addition, by definition $\gamma_{j}=\mathbb{I}(\beta_{j}\neq0)$, and
the term $\Vert \gamma\Vert _{1}$ can further be seen as an $l_{0}$ norm on
$\beta$, as $\Vert \gamma\Vert _{1}=\sum_{j=1}^{p}|\mathbb{I}(\beta_{j}\neq
0)|=\lim_{s\rightarrow0}\sum_{j=1}^{p}\beta_{j}^{s}$, which is the
$l_{0}$ norm by definition. Here, we have used the assumption that
$0^{0}=0$. We can express the fourth term in~(\ref{method1}) by $\rho
_{\lambda,\kappa,\sigma^{2}}\Vert \beta\Vert _{0}$.

%s3.2 ###
\subsection{Parameter estimation}\label{sec32}
\begin{figure}

\includegraphics{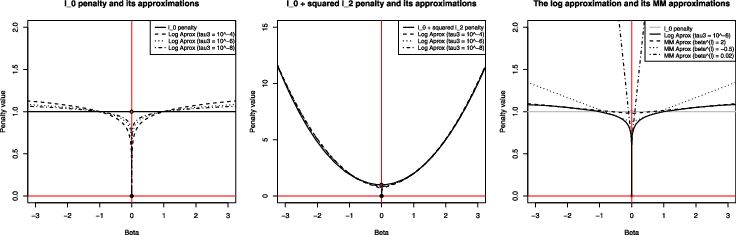}

\caption{The penalty functions and related approximations.}
\label{logsumfigure1}
\end{figure}

Now given all other parameters fixed, the MAP estimator of $\sigma^{2}$
can be derived by first making a derivative of (\ref{method1}) with
respect to $\sigma^{2}$, setting the derivative to zero, and then
solving the equation for $\sigma^{2}$. The estimation of $\beta$ is
further carried out given $\sigma^{2}$ is fixed. With fixed $\sigma
^{2}$ and the interpretations of regularization estimation given above,
(\ref{method1}) has an equivalent representation given by
%
%e6 ###
\begin{equation}
\label{6}
L(\beta;\lambda,\rho_{\lambda,\kappa,\sigma^{2}})=\Vert y-X\beta
\Vert _{2}^{2}+\lambda\Vert \beta\Vert _{2}^{2} + \rho_{\lambda,\kappa,\sigma
^{2}}\Vert \beta\Vert _{0}+\mathrm{const}.
\end{equation}
Note that here we have multiplied (\ref{method1}) with $\sigma^{2}$.
Now with (\ref{6}), we can construct an iteration scheme to obtain (\ref{5}). At the $(m+1)$th iteration, the iteration scheme is given by
%
%e7 ###
\begin{eqnarray}
\label{7}
(\widehat{\sigma}^{2})^{(m+1)}&=& \frac{\sum_{i=1}^{n}(y_{i}-\sum
_{j=1}^{p}x_{ij}\widehat{\gamma}_{j}^{(m)}\widehat{\beta
}_{j}^{(m)})^{2} + \lambda\sum_{j=1}^{p}\widehat{\gamma
}_{j}^{(m)}(\widehat{\beta}_{j}^{(m)})^{2}+2\tau_{2}}{n+\sum
_{j=1}^{p}\widehat{\gamma}_{j}^{(m)}+2\tau_{1}+2},\nonumber\hspace*{-35pt}\\[-2pt]
\widehat{\beta}^{(m+1)} &=& \arg\min_{\beta} L\bigl(\beta;\lambda,\rho
_{\lambda,\kappa,(\widehat{\sigma}^{2})^{(m+1)}}\bigr),\hspace*{-35pt}\\[-2pt]
\widehat{\gamma}^{(m+1)} &=&\bigl(\mathbb{I}\bigl(\widehat{\beta}_{1}^{(m+1)}\neq
0\bigr),\mathbb{I}\bigl(\widehat{\beta}_{2}^{(m+1)}\neq0\bigr),\ldots,\mathbb
{I}\bigl(\widehat{\beta}_{p}^{(m+1)}\neq0\bigr)\bigr).\nonumber\hspace*{-35pt}
\end{eqnarray}
Note that the objective function (\ref{6}) involves an $l_{0}$ norm,
which by definition, is not continuous. Therefore, related optimization
tasks in the second term of (\ref{7}) require some refinements. Here,
we adopt a relaxation approach to tackling the optimization problem. We
begins the approach by noting that, mathematically the~$l_{0}$ norm on
a $p$-dimensional vector $\beta$ can be expressed as
%
%e8 ###
\begin{equation}
\label{9}
\Vert \beta\Vert _{0} = \lim_{\tau_{3}\rightarrow0}\sum_{j=1}^{p}\frac{\log
(1+\tau_{3}^{-1}|\beta_{j}|)}{\log(1+\tau_{3}^{-1})},
\end{equation}
which can be verified by seeing (\ref{9}) as a function of $\tau_{3}$
and using l'H\^{o}pital's rule. A more detailed discussion on the
properties of the log-sum function on the right-hand side of (\ref{9})
is given in Supplementary Material~\cite{sup}. With representation~(\ref{9}), the
objective function~(\ref{6}) can be reexpressed as
%
%e9 ###
\begin{eqnarray}
\label{10}
L(\beta;\lambda,\rho_{\lambda,\kappa,\sigma^{2}})&=&\Vert y-X\beta
\Vert _{2}^{2}+\lambda\Vert \beta\Vert _{2}^{2}
\nonumber
\\[-9pt]
\\[-9pt]
\nonumber
& & {}+ \rho_{\lambda,\kappa,\sigma^{2}}\Biggl\{\lim_{\tau_{3}\rightarrow
0}\sum_{j=1}^{p}\frac{\log(1 + \tau_{3}^{-1}|\beta_{j}|)}{\log(1+\tau
_{3}^{-1})}\Biggr\}+\mathrm{const}.
\end{eqnarray}
If $\tau_{3}$ is small enough, the log-sum function on the right-hand
side of (\ref{10}) will give an approximate representation of $\Vert \beta
\Vert _{0}$. Graphical representations for the log-sum function with
different $\tau_{3}$ and their mixtures with the squared $l_{2}$ norm
can be found in the left and middle panels of Figure~\ref{logsumfigure1}.
In addition, since the log-sum function in (\ref{10})
is continuous in $\beta$, the combinatorial nature of $\Vert \beta\Vert _{0}$
is relaxed. However, the term $\log(1+\tau_{3}^{-1}|\beta_{j}|)$ is not
convex in $\beta_{j}$, and replacing $\Vert \beta\Vert _{0}$ with~(\ref{9}) in
(\ref{6}) still makes objective function (\ref{10}) remain nonconvex.
To tackle this problem, a majorization--minimization algorithm is
adopted. Majorization-minimization (MM) algorithms~\cite{hunterandlange05,wuandlange08} are a set of analytic procedures aiming
to tackle difficult optimization problems by modifying their objective
functions so that solution spaces of the modified ones are easier to
explore. For an objective function $g(\theta)$, the modification
procedure relies on finding a function $h(\theta;\theta^{(l)})$
satisfying the following properties:
%
%e10 ###
\begin{eqnarray}
\label{101}
h\bigl(\theta; \theta^{(l)}\bigr)&\geq& g(\theta)  \qquad  \mbox{for all }\theta,
\nonumber
\\[-9pt]
\\[-9pt]
\nonumber
h\bigl(\theta^{(l)}; \theta^{(l)}\bigr)&=&g\bigl(\theta^{(l)}\bigr).
\end{eqnarray}
In (\ref{101}), the objective function $g(\theta)$ is said to be
majorized by $h(\theta;\theta^{(l)})$. In this sense, $h(\theta;\theta
^{(l)})$ is called the majorization function. In addition, (\ref{101})
implies that $h(\theta; \theta^{(l)})$ is tangent to $g(\theta)$ at
$\theta^{(l)}$. Moreover, if $\theta^{(l+1)}$ is a minimizer of
$h(\theta; \theta^{(l)})$, then (\ref{101}) further implies that
%
%e11 ###
\begin{equation}
\label{104}
g\bigl(\theta^{(l)}\bigr)=h\bigl(\theta^{(l)}; \theta^{(l)}\bigr)\geq h\bigl(\theta^{(l+1)};
\theta^{(l)}\bigr)\geq g\bigl(\theta^{(l+1)}\bigr),
\end{equation}
which means that the iteration procedure $\theta^{(l)}$ pushes $g(\theta
)$ toward its minimum.

Now we turn back to the function on the right-hand side of (\ref{10}).
Note that, since $\log(\theta)$ is a concave function of $\theta$ for
$\theta>0$, therefore the inequality
%
%e12 ###
\begin{equation}
\label{102}
\log(\theta^{\prime}) + \frac{\theta}{\theta^{\prime}}-1\geq\log(\theta)
\end{equation}
holds for all $\theta>0$ and $\theta^{\prime}>0$. Note that the
left-hand side of (\ref{102}) is convex in~$\theta$. In addition, if
we let $\theta^{\prime}=\theta$, then (\ref{102}) becomes an equality,
which implies that the left-hand side of (\ref{102}) satisfies the
properties stated in (\ref{101}), therefore is a valid function for
majorizing $\log(\theta)$.
\begin{proposition}
\label{proposition1}
Define $\rho_{\tau_{3}}=1/\log(1+\tau_{3}^{-1})$ and let $L^{\prime
}(\beta;\lambda,\rho_{\lambda,\kappa,\sigma^{2}})$ be the same as (\ref{6})
but without the constant term. Then $L^{\prime}(\beta;\lambda,\rho
_{\lambda,\kappa,\sigma^{2}})$ can be majorized by the following function:
%
%e13 ###
\begin{equation}
\label{mmfunction1}
L^{\prime\prime}(\beta;\lambda,\rho_{\lambda,\kappa,\sigma^{2}},\beta
^{\prime}) = \Vert y-X\beta\Vert _{2}^{2}+\lambda\Vert \beta\Vert _{2}^{2} + \rho
_{\lambda,\kappa,\sigma^{2}}h_{2}(\beta;\beta^{\prime}),
\end{equation}
where
%
%e14 ###
\begin{equation}
\label{11}
h_{2}(\beta;\beta^{\prime})= \lim_{\tau_{3}\rightarrow0}\rho_{\tau
_{3}}\sum_{j=1}^{p}\biggl(\log(1 + \tau_{3}^{-1}|\beta
_{j}^{\prime}|) + \frac{|\beta_{j}|+\tau_{3}}{|\beta
_{j}^{\prime}|+\tau_{3}}-1\biggr).
\end{equation}
\end{proposition}
\begin{pf}
Let $h_{1}(\beta
)=\Vert y-X\beta\Vert _{2}^{2}+\lambda\Vert \beta\Vert _{2}^{2}$. Assume $\beta
^{(l+1)}$ minimizes $L^{\prime\prime}(\beta;\break\lambda, \rho_{\lambda,\kappa
,\sigma^{2}},\beta^{\prime})$ given $\beta^{\prime}=\beta^{(l)}$. Then
with (\ref{9}) and the inequality (\ref{102}), the quantity $L^{\prime
}(\beta^{(l+1)};\lambda,\rho_{\lambda,\kappa,\sigma^{2}})$ can be
bounded in a way such that
%
%e15 ###
\begin{eqnarray}
\label{111}
L^{\prime}\bigl(\beta^{(l+1)};\lambda,\rho_{\lambda,\kappa,\sigma
^{2}}\bigr)&=&h_{1}\bigl(\beta^{(l+1)}\bigr)+\rho_{\lambda,\kappa,\sigma^{2}}\lim_{\tau
_{3}\rightarrow0}\rho_{\tau_{3}}\sum_{j=1}^{p}\log\bigl(1 + \tau
_{3}^{-1}\bigl|\beta_{j}^{(l+1)}\bigr|\bigr)\nonumber\\
&\leq& h_{1}\bigl(\beta^{(l+1)}\bigr)+\rho_{\lambda,\kappa,\sigma^{2}}h_{2}\bigl(\beta
^{(l+1)};\beta^{(l)}\bigr)\\
&=&L^{\prime\prime}\bigl(\beta^{(l+1)};\lambda,\rho_{\lambda,\kappa,\sigma
^{2}},\beta^{(l)}\bigr)\nonumber
\end{eqnarray}
which verifies the first condition stated in (\ref{101}). For $\beta=
\beta^{\prime}$, $h_{2}(\beta;\beta^{\prime})$ is equal to the log-sum
function in (\ref{9}), which verifies the second condition stated in
(\ref{101}) and completes the proof.
\end{pf}

A graphical representation of using MM algorithms in approximating the
log-sum function in (\ref{9}) can be found in the right panel of Figure
\ref{logsumfigure1}. From the argument given above, we can construct\vadjust{\goodbreak}
an iteration scheme to obtain the minimizer of $L(\beta;\lambda,\rho
_{\lambda,\kappa,\sigma^{2}})$, with the $l_{0}$ norm, or equivalently
the log-sum function, replaced by $h_{2}(\beta;\beta^{\prime})$ defined
in Proposition~\ref{proposition1}. For example, in (\ref{7}), $\widehat
{\beta}^{(m+1)}$ can be obtained by carrying out the following
iteration scheme:
%
%e16 ###
\begin{eqnarray}
\label{12}
\qquad &&\widehat{\beta}^{(m+1,l+1)}
\nonumber
\\[-4pt]
\\[-12pt]
\nonumber
&&\qquad=\arg\min_{\beta}\Biggl\{\Vert y-X\beta\Vert _{2}^{2}
+ \lambda\Vert \beta\Vert _{2}^{2} + \rho_{\lambda,\kappa,(\widehat{\sigma
}^{2})^{(m+1)}}\sum_{j=1}^{p}\widehat{\phi}_{j}^{(m+1,l)}|\beta
_{j}|\Biggr\}
\end{eqnarray}
over index\vspace*{2pt} $l$, where $\widehat{\phi}_{j}^{(m+1,l)}=\lim_{\tau
_{3}\rightarrow0}[\log(1+\tau_{3}^{-1})(|\widehat{\beta
}_{j}^{(m+1,l)}|+\tau_{3})]^{-1}$. The procedure of using iteration
scheme (\ref{12}) in obtaining the minimizer for the objective function
(\ref{6}), or equivalent (\ref{10}), is called the BAVA-MIO (BAyesian
VAriable selection using a Majorization--mInimization apprOach), and the
resulting minimizer is called the BAVA-MIO estimator.

%
%f1 ###

Note that the last term on the right-hand side of (\ref{12}) is a
linear combination of $\widehat{\phi}_{j}^{(m+1,l)}|\beta_{j}|$, a
convex function of $\beta_{j}$, therefore given $\Vert y-X\beta\Vert _{2}^{2} +
\lambda\Vert \beta\Vert _{2}^{2}$ is convex in $\beta$, the whole objective
function in (\ref{12}) will be
convex in $\beta$, which guarantees that the iteration scheme will
converge. In addition, the minimizer (\ref{12}) can be obtained by
using the coordinate descent algorithm proposed by Friedman et al.~\cite{friedmanetal07}. In practice, the coordinate descent algorithm is
based on iteratively cycling a one-dimensional soft-thresholding
scheme. Given that $\rho_{\lambda,\kappa,\sigma^{2}}$ is fixed, at the
$(m_{1}+1)$th iteration, the soft-thresholding scheme for the $j$th
coordinate is given by
%
%e17 ###
\begin{equation}
\label{131}\quad
\tilde{\beta}_{j}^{(m_{1}+1)} = \Biggl(\sum_{i=1}^{n}x_{ij}^{2}+\lambda
\Biggr)^{-1}\operatorname{ST}\Biggl(\sum_{i=1}^{n}x_{ij}\tilde{r}_{i,-j}^{(m_{1})},\rho
_{\lambda,\kappa,\sigma^{2}}\frac{\tilde{\phi}_{j}^{(m_{1})}}{2}\Biggr),
\end{equation}
where $\tilde{r}_{i,-j}^{(m_{1})}=y_{i}-\sum_{j^{\prime}\neq
j}x_{ij^{\prime}}\tilde{\beta}_{j^{\prime}}^{(m_{1}^{\prime})}$, with
$m_{1}^{\prime}=m_{1}+1$ for $j^{\prime}=1,2,\ldots,j-1$, and
$m_{1}^{\prime}=m_{1}$ for $j^{\prime}=j+1,j+2,\ldots,p$, and $\tilde
{\phi}_{j}^{(m_{1})}=\lim_{\tau_{3}\rightarrow0}[\log(1+\tau
_{3}^{-1})(|\tilde{\beta}_{j}^{(m_{1})}|+\tau_{3})]^{-1}$. Here
$\operatorname{ST}(a,b)$ is a soft-thresholding operator defined by $\operatorname{ST}(a,b)=\operatorname{
sign}(a)(|a|-b)_{+}$. A detailed derivation of (\ref{131}) is given in
Appendix A of Supplementary Material~\cite{sup}.

%s3.3 ###
\subsection{Choosing hyperparameters}\label{sec33}

Choosing appropriate hyperparameters for prior construction is an
important issue in many Bayesian inference problems. For
hyperparameters present in the model (\ref{2}), we consider the triple
$(\lambda,\tau_{1},\tau_{2})$ first. One principle we adopt in
parameterizing the hyperparameters is that as the number of samples $n$
increases, the impact of the hyperparameters in parameter estimation
will become less significant. In addition, we let $\tau_{1}=\tau_{2} +
1$, so that the prior expectation of $\sigma^{2}$ is equal to 1. Given
these conditions, one of the possible choices is $(\lambda,\tau_{1},\tau
_{2})=(1/\sqrt{n},p\log p/\sqrt{n}+1, p\log p/\sqrt{n})$. We will
discuss other possible settings in the simulation study in the later section.

Now we consider the prior inclusion probability $\kappa$. In some
circumstances, data-driven empirical Bayes approaches~\cite{georgeandfoster00} are proposed to obtain $\kappa$, while in other
circumstances full Bayesian methods that assign priors on $\kappa$ are
proposed. For example, please see~\cite{liangetal07}. Unlike previously
proposed approaches, in which single point estimates were obtained for
$\kappa$, we adopt an approach by specifying a feasible region for the function
%
%e18 ###
\begin{equation}
\label{chos1}
\psi(\kappa)= \tfrac{1}{2}\bigl[\sigma^{2}\log\bigl(2\pi\sigma^{2}\lambda
^{-1}(1-\kappa)^{2}/\kappa^{2}\bigr)\bigr]\widehat{\phi}^{(0)},
\end{equation}
and carry out parameter estimation under different values of $\psi
(\kappa)$. Here we have assumed $\widehat{\phi}_{j}^{(0)}=\widehat{\phi
}^{(0)}$ for $j=1,2,\ldots,p$. Note that\vspace*{-2pt} by definition the term
$\widehat{\phi}_{j}^{(0)}$ is a function of the initial value $\widehat
{\beta}_{j}^{(0)}$. The function $\psi(\kappa)$ is the threshold used
in the soft-thresholding scheme (\ref{131}). We carry out the
parameter estimation with values in the feasible region and look for
which values of $\psi(\kappa)$ lead to the best performance measured by
criteria such as ten fold cross validation or the Bayes factor. Under
this approach, estimated parameters can be seen as functions of $\kappa
$ on the feasible region. Given different values of $\kappa$,
curve-like paths for estimated parameters can be obtained. The main
reason we adopt this ``whole-path'' fitting strategy is that the
optimization procedure may get stuck in some stationary points. It can\vspace*{-1.5pt}
occur in a situation in which we need an initial value $\widehat{\phi
}_{j}^{(1)}$ to run the iteration scheme (\ref{12}). By definition,
$\widehat{\phi}_{j}^{(1)}$ is a function of $\widehat{\beta
}_{j}^{(1)}$, which by definition,\vspace*{2pt} is a function of $\psi(\kappa)$. As
pointed out by Cand\'{e}s et al.~\cite{candesetal08} and Mazumder et
al.~\cite{mazumderetal09}, different $\widehat{\phi}_{j}^{(1)}$ may
lead to different solutions for the minimizer. Under this situation, a
global minimum may not be guaranteed. By using the strategy given
above, we can run the iteration scheme (\ref{12}) with a large number
of possible values of~$\widehat{\phi}_{j}^{(1)}$,\vspace*{2pt} therefore eliminating
the possibility that the solution is stuck in some local minima.

Our approach is similar to the one using a fixed grid on the tuning
parameter and then running parameter estimation with different values
of the tuning parameter. This fixed grid approach to tuning parameter
selection has been adopted in~\cite{genkinetal07,friedmanetal07,meieretal08} and is
advocated by~\cite{wuandlange08,friedmanetal09} for fast and accurate parameter estimation.

%s3.4 ###
\subsection{A toy example}\label{sec34}
Here, we provide a toy example to illustrate the BAVA-MIO estimation.
We let the number of samples $n=100$ and the number of covariates
$p=1\mbox{,}000$. For regression coefficients $\beta=(\beta_{1},\beta
_{2},\ldots,\beta_{1\mbox{,}000})$, we let $\beta_{250}= 2$, $\beta
_{500}=-3.2$, $\beta_{750}=-1.25$, $\beta_{1\mbox{,}000}=5.44$, and $\beta
_{j}=0$ for all $j$'s $\in\{1,2,\ldots,1\mbox{,}000\}\setminus\{
250,500,750,1\mbox{,}000\}$. We generate each row of $X$ independently
identically from $\operatorname{MVN}(0,I_{p\times p})$, and then calculating $Y=X\beta
+\varepsilon$ with $\varepsilon\sim$ $\operatorname{MVN}(0,I_{n\times n})$. For the
hyperparameters, we let $\tau_{1}= 0.2p\log(p)/\sqrt{n} + 1$, $\tau
_{2}= 0.2p\log(p)/\sqrt{n}$ and $\lambda= 1/\sqrt{n}$. Further let
$\tau_{3}=10^{-6}$. We use 100 equal spaced points to form a grid for
$\Psi(\kappa)$. We perform two BAVA-MIO estimations: one uses the Bayes
factor and the other uses ten fold cross validation for tuning
parameter selection. Remember the index set $S$ is defined by $S=\{
j\dvtx \gamma_{j}=1\}$. We define the Bayes factor between models $\mathcal
{M}_{S^{\prime}}$ and $\mathcal{M}_{S}$ by
%
%e19 ###
\begin{equation}
\label{bayesfactor}
\mathrm{BF}(\mathcal{M}_{S^{\prime}},\mathcal{M}_{S};y) = \frac{f(y| \gamma^{\prime},\tau_{1},\tau_{2},\kappa,\lambda)}{f(y| \gamma,\tau
_{1},\tau_{2},\kappa,\lambda)},
\end{equation}
where the term $f(y| \gamma,\tau_{1},\tau_{2},\kappa,\lambda)$ refers
to the marginalized likelihood with $\beta$ and $\sigma^{2}$ being
integrated out with respect to their prior probability measures. For
the Bayesian model stated in (\ref{2}), the marginalized likelihood has
a closed form representation given by
\begin{eqnarray*}
\label{19}
f(y| \gamma,\tau_{1},\tau_{2},\kappa,\lambda)&=&\frac{\pi
^{-n/2}}{|\lambda^{-1}X_{S}^{T}X_{S}+ I_{\gamma}|^{1/2}}\frac{(2\tau
_{2})^{\tau_{1}}}{\Gamma(\tau_{1})}\Gamma\biggl(\frac{n+2\tau
_{1}}{2}\biggr)\\
& &{}\times\bigl(y^{T}(\lambda^{-1}X_{S}X_{S}^{T}+I_{n})^{-1}y
+2\tau_{2}\bigr)^{-[(n+2\tau_{1})/2]}.
\end{eqnarray*}
In subsequent sections we will use the measure (\ref{bayesfactor}) for
variable selection. In addition, for all variable selection tasks using
(\ref{bayesfactor}), the baseline model $\mathcal{M}_{S}$ will always
refer to the null model.

\begin{figure}

\includegraphics{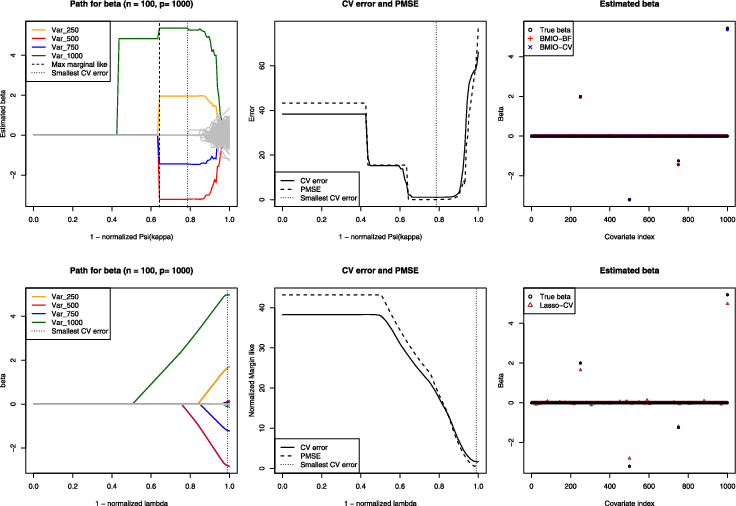}

\caption{Estimation results for the toy example.}
\label{toyfigure1}
\end{figure}

The results are shown in Figure~\ref{toyfigure1}. The path plot in the
top left panel of Figure~\ref{toyfigure1} shows that nonzero
coefficients entered into the model earlier under the BAVA-MIO
estimation. In addition, the paths of estimated coefficients behave
similar to those under the hard-thresholding estimation, that is, once
a coefficient is estimated to be nonzero, the corresponding estimation
path makes a sharp jump to the nonthresholded value. Moreover, due to
the presence of the squared $l_{2}$ norm in the objective function, the
number of selected covariates can be larger than the number of samples.
Throughout the estimation procedure, the maximum number of selected
covariates is 831, which is much larger than the number of samples
$n=100$. Here we also provide the lasso estimation for regression
fitting with the same data. The results are shown in the bottom panel
of Figure~\ref{toyfigure1}. As compared with the lasso estimation, in
which 33 covariates are selected using ten fold cross validation, the
BAVA-MIO estimations using the Bayes factor and ten fold cross
validation correctly select covariates with nonzero coefficients. In
addition, as shown in the right panel of Figure~\ref{toyfigure1},
values of the nonzero coefficients are also estimated more accurately
under the BAVA-MIO estimations.
%
%f2 ###

%s4 ###
\section{Simulation studies}\label{sec4}
In this section, we conduct two simulation studies. The first one is a
general assessment on the performance of the BAVA-MIO estimation. The
second one focuses the performance of the BAVA-MIO estimation under
various situations in which the Irrepresentable Condition may or may
not hold.

%s4.1 ###
\subsection{Simulation study I}\label{sec41}
In the first simulation study, we compare the BAVA-MIO estimation with
other estimation approaches by fitting regression model $Y=X\beta
+\varepsilon$ with data generated from different simulation schemes. Here
$Y$ and $\varepsilon$ are $n$-dimensional vectors, $X$ is an $n\times p$
matrix and $\beta$ is a $p$-dimensional vector. We assume each entry
in $\varepsilon$ is i.i.d. from $\operatorname{Normal}(0,\sigma_{Y}^{2})$, and each row
in the design matrix $X$ is i.i.d. from $\operatorname{MVN}(0,\Sigma_{X})$. Throughout
the whole simulation study, we let $p=120$. For regression coefficient
vector $\beta=(\beta_{1},\beta_{2},\ldots,\beta_{p})$, we generate
$\beta_{j}$ from $\operatorname{Normal}(0,1)$ for $j=1,2,\ldots,10$ and let $\beta
_{j}=0$ for $j=11,12,\ldots,120$. That is, we have 10 nonzero and 110
zero coefficients in the ``true'' model. In addition, we use different
values of $(\Sigma_{X},\sigma_{Y}^{2},n)$ in generating the design
matrix $X$ and the error term $\varepsilon$. We apply three different
$\Sigma_{X}$ to generate the design matrix. The first one has an
independent structure with diagonal terms equal to 1 and off-diagonal
terms equal to 0. The second one has a covariance structure such that
$(\Sigma_{X})_{ij}=1$ for $i=j$ and $(\Sigma_{X})_{ij}=0.5$ for $i\neq
j$. The third one has a covariance structure such that $(\Sigma
_{X})_{ij}=0.5^{|i-j|}$. Now define the signal-to-noise ratio by SNR $=
\sqrt{\mathbb{E}(\beta^{T}\Sigma_{X}\beta)/\sigma_{Y}^{2}}$. We
consider $\sigma_{Y}^{2}=10, 1$ and $0.2$ in generating the error term
$\varepsilon$. For $\Sigma_{X}=I_{10\times10}$, these values correspond
to SNR $= 1, 3.16$ and $7.07$, respectively. For practical purposes, we
will use the labels SNR $ = 1$ for experiments using $\sigma
_{Y}^{2}=10$, SNR $ = 3.16$ for experiments using $\sigma_{Y}^{2}=1$
and SNR $=7.07$ for experiments using $\sigma_{Y}^{2}=0.2$. By using
$X$, $\beta$ and $\varepsilon$, the response vector $Y$ is calculated by
$Y=X\beta+\varepsilon$. For the number of samples, we consider five values
$n=40, 80,120,160$ and $200$. With three different structures for
$\Sigma_{X}$, three different values for $\sigma_{Y}^{2}$, and five
different values for $n$, we have total $3\times3\times5=45$
simulation experiments.

Here we describe hyperparameter settings in BAVA-MIO estimations. We
let hyperparameters $(\lambda,\tau_{1},\tau_{2})=(1/\sqrt{n},p\log
p/\sqrt{n}+1, p\log p/\sqrt{n})$ for the cases of SNR $=3.16$ and
$7.07$. For the case of SNR $=1$, we use
%
%e20 ###
\begin{eqnarray}
\label{simu1}
\lambda&=&\biggl(\frac{1-\widehat{\operatorname{corr}}}{\widehat{\operatorname
{corr}}}\biggr)^{2}\sqrt{\frac{p}{n}}\log p,\nonumber\\
\tau_{1}&=&\biggl(\frac{\widehat{\operatorname{corr}}}{1-\widehat{\operatorname
{corr}}}\biggr)\biggl(\frac{p\log p}{\sqrt{n}}\biggr)^{\widehat{\operatorname
{corr}}/\log n}+1,\\
\tau_{2}&=&\frac{1}{\sqrt{n}}\biggl(\frac{p\log p}{\sqrt{n}}\biggr)^{1 +
\widehat{\operatorname{corr}}/\log n},\nonumber
\end{eqnarray}
where $\widehat{\operatorname{corr}}$ is an average over the top 10 percent
absolute values of the sample correlations between response $Y$ and
covariates $X$. For tuning parameter selection, we use two criteria:
the Bayes factor, which is defined in (\ref{bayesfactor}), and ten-fold cross validation. The resulting estimators are called BMIO-BF and
BMIO-CV, respectively.

We also carry out three other estimation approaches for comparisons.
The first one is the lasso~\cite{tibshirani96}. We use R package
``glmnet'' to obtain the lasso estimates. The tuning parameter is
selected using ten fold cross validation. The second approach is the
relaxed lasso~\cite{meinshausen07}. We use R package ``relaxo,'' which
is the companion software to~\cite{meinshausen07}, to obtain the
relaxed lasso estimates. The tuning parameter is selected using ten
fold cross validation with 100 values of scaling parameters equally
spaced in $[0,1]$. The third approach is the adaptive lasso~\cite{zou06}. We use R package ``parcor'' to obtain the adaptive lasso
estimates with the default setting that uses the lasso estimate as the
initial value for the weight and selects the tuning parameter $\lambda$
via ten fold cross validation.

We collect several performance measures at each simulation run. The
first one is the standardized $l_{2}$ distance between a given estimate
$\widehat{\beta}$ and the true regression coefficient vector $\beta$,
which is defined by
\[
l_{2}\mathrm{\mbox{-}dis}(\widehat{\beta})=\sqrt{\frac{\sum_{j=1}^{p}(\widehat
{\beta}_{j}-\beta_{j})^{2}}{\sum_{j=1}^{p}\beta_{j}^{2}}}.
\]
The second one is the predictive mean squared error of $\widehat{\beta
}$ for a test data set, which is defined by
\[
\operatorname{PMSE}(\widehat{\beta}) = \frac{\sum_{i=1}^{n_{\mathrm
{test}}}(x_{i,\mathrm{test}}^{T}\widehat{\beta}-x_{i,\mathrm{test}}^{T}\beta
)^{2}}{n_{\mathrm{test}}}.
\]
The test data set contains $n_{\mathrm{test}}=n\times10$ data points
generated using a simulation scheme the same as the training data set.
The third
one is the number of coefficients with nonzero estimated values
$|\widehat{S}|$, where $\widehat{S}=\{j\dvtx \widehat{\beta}_{j}\neq0\}$.
The final one is the sign function-based false positive rates, which is
defined by
\[
\mathrm{S\mbox{-}FPR}=\frac{\#\{j\in\widehat{S}\dvtx \operatorname{sign}(\widehat{\beta
}_{j})\neq\operatorname{sign}(\beta_{\mathrm{true},j})\}}{|\widehat{S}|},
\]
where the sign function $\operatorname{sign}(\cdot)$ is defined in Section~\ref{sec2}.

%f3 ###
\begin{figure}

\includegraphics{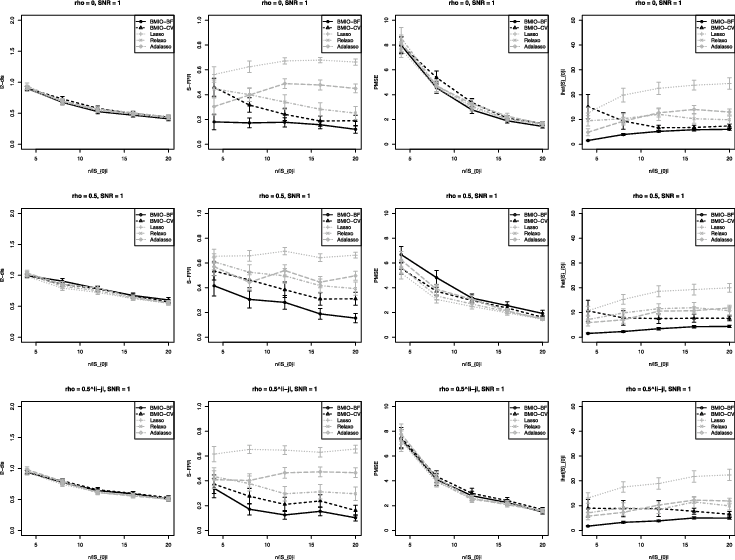}

\caption{Simulation results given SNR $= 1$. Top: Model 1 (covariance
matrix with off-diagonal terms equal to $0$); Middle: Model 2
(covariance matrix with off-diagonal terms equal to $0.5$); Bottom:
Model~3 (covariance matrix with off-diagonal terms following a
specified covariance structure). First column: standardized
$l_{2}$-distance between estimated and true values; Second column: sign
function-adjusted false positive rate; Third column: prediction mean
squared error; Fourth column: number of nonzero estimates.}
\label{simuIfig1}
\end{figure}

\begin{figure}

\includegraphics{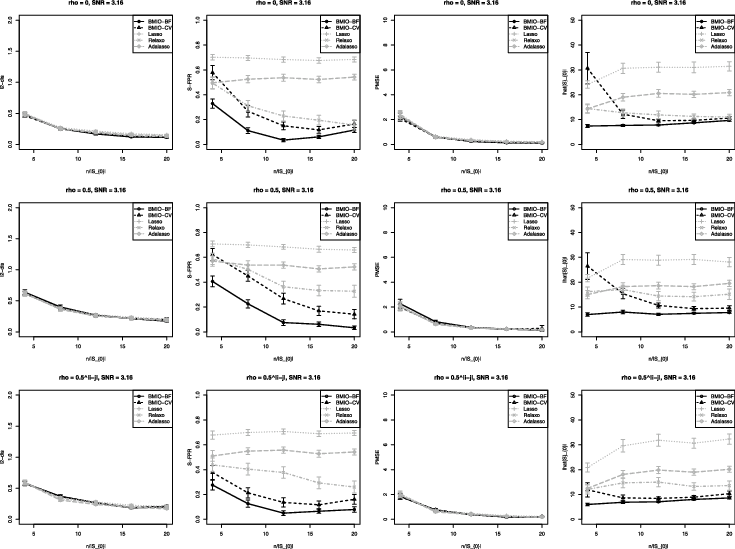}

\caption{Simulation results given SNR $= 3.16$. Top: Model 1
(covariance matrix with off-diagonal terms equal to $0$); Middle: Model
2 (covariance matrix with off-diagonal terms equal to $0.5$); Bottom:
Model 3 (covariance matrix with off-diagonal terms following a
specified covariance structure). First column: standardized
$l_{2}$-distance between estimated and true values; Second column: sign
function-adjusted false positive rate; Third column: prediction mean
squared error; Fourth column: number of nonzero estimates.}
\label{simuIfig2}
\end{figure}

\begin{figure}

\includegraphics{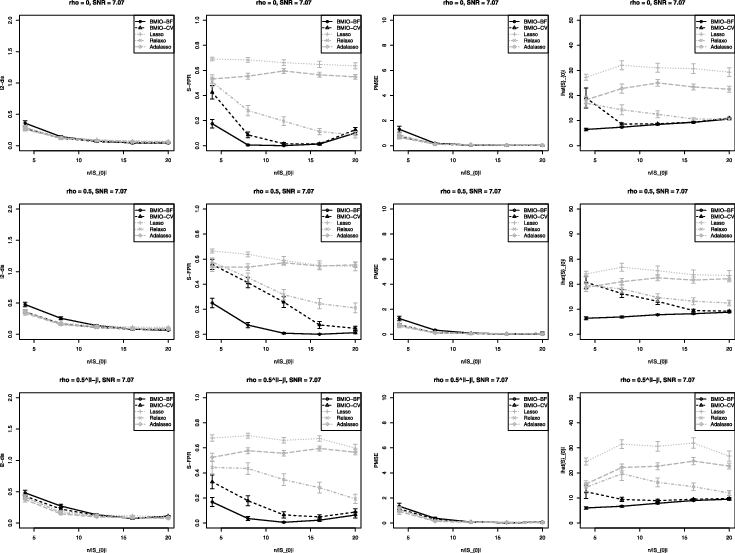}

\caption{Simulation results given SNR $= 7.07$. Top: Model 1
(covariance matrix with off-diagonal terms equal to $0$); Middle: Model
2 (covariance matrix with off-diagonal terms equal to $0.5$); Bottom:
Model 3 (covariance matrix with off-diagonal terms following a
specified covariance structure). First column: standardized
$l_{2}$-distance between estimated and true values; Second column: sign
function-adjusted false positive rate; Third column: prediction mean
squared error; Fourth column: number of nonzero estimates.}
\label{simuIfig3}
\end{figure}

For each of the 45 simulation experiments, we generate 100 runs to
collect the four performance measures. We then plot the average of each
performance measure against the ratio $n/|S|$, that is, the ratio
between the number of samples and the number of true coefficients with
nonzero values. These plots are shown in Figures~\ref{simuIfig1},~\ref{simuIfig2} and~\ref{simuIfig3} for SNR $= 1, 3.16$ and $7.07$,
respectively. From the three figures, we can see none of the estimation
approaches can dominate the others in all four performance measures. In
most cases, BAVA-MIO based estimations have smaller sign function-based
false positive rates, as shown in the second column of each figure. It
implies that more accurate variable selection may be done using the
BAVA-MIO estimations. These findings become more significant as the
number of samples increases. In addition, BAVA-MIO estimations have
fewer numbers of nonzero estimates, as shown in the fourth column of
each figure. Moreover, since the BAVA-MIO estimation using the Bayes
factor has relatively fewer numbers of nonzero estimates, it is
surprising that the PMSE and $l_{2}$-dis measures under the BMIO-BF
estimation are comparable to those under other estimation approaches,
for example, in the cases with SNR $= 3.16$ and in some cases with SNR
$=1$. However, we also noticed that the BMIO-BF estimation has higher
values in the PMSE and $l_{2}$-dis in the cases with SNR $=7.07$,
particularly in the situations in which the number of samples is small.
%

%
%f4 ###

%
%f5 ###

%s4.2 ###
\subsection{Simulation study II}\label{sec42}

In the second simulation study, we investigate the impact of the
Irrepresentable Condition on the performance of BAVA-MIO estimation in
variable selection. Before stating the Irrepresentable Condition, we
give some notation definitions. We define $S_{0}=\{j\dvtx \beta_{j}\neq
0,\mbox{ for some}\break j\in\{1,2,\ldots,p\}$ and $S_{0}^{c}=\{1,2,\ldots,p\}
\setminus S_{0}$. Let $\beta_{S_{0}}$ denote the coefficients with
indices in $S_{0}$ and $\beta_{S_{0}^{c}}$ the coefficients with
indices in $S_{0}^{c}$. Similar definitions are also applied to
$X_{S_{0}}$ and $X_{S_{0}^{c}}$. An estimator $\widehat{\beta}(n)$ is
said to be sign consistent in estimating $\beta$ if the probability of
the event $\{\operatorname{sign}(\widehat{\beta}(n))=\operatorname{ sign}(\beta)\}$
approaches to $1$ as $n\rightarrow\infty$. Given the sign consistency
holds, the estimated index set $\widehat{S}_{0}=\{j\dvtx \widehat{\beta
}_{j}\neq0\}$ will be the same as the true index set $S_{0}$,
therefore the sign consistency implies variable selection consistency,
that is, asymptotically with probability one, nonzero-valued
coefficients will have nonzero estimated values and zero-valued
coefficients will be estimated with zero values.

Zhao and Yu~\cite{zhaoandyu06} showed that if one wants the lasso
estimation to achieve the sign consistency, then the design matrices
$X$ must satisfy the following condition:
%
%e21 ###
\begin{equation}
\label{irr1}
\Vert X_{S_{0}^{c}}^{T}X_{S_{0}}(X_{S_{0}}^{T}X_{S_{0}})^{-1}\operatorname
{sign}(\beta_{S_{0}})\Vert _{\infty}<1,
\end{equation}
where $\beta_{S_{0}}$ is the vector of nonzero-valued coefficients.
The condition (\ref{irr1}) is called the (Weak) Irrepresentable
Condition. If the Irrepresentable Condition (\ref{irr1}) fails to
hold, then the sign consistency will never occur even when $n\rightarrow
\infty$. An intuitive way to explain the Irrepresentable Condition is
to see the quantity
$X_{S_{0}^{c}}^{T}X_{S_{0}}(X_{S_{0}}^{T}X_{S_{0}})^{-1}$ as a least
squares estimate for the regression $X_{S_{0}^{c}}$ on $X_{S_{0}}$. In
this sense, the Irrepresentable Condition states that the largest
amount of coefficients for the regression $X_{S_{0}^{c}}$ on
$X_{S_{0}}$ should not exceed 1, that is, $X_{S_{0}^{c}}$ is
``irrepresentable'' in terms of $X_{S_{0}}$.

%
%f6 ###
\begin{figure}

\includegraphics{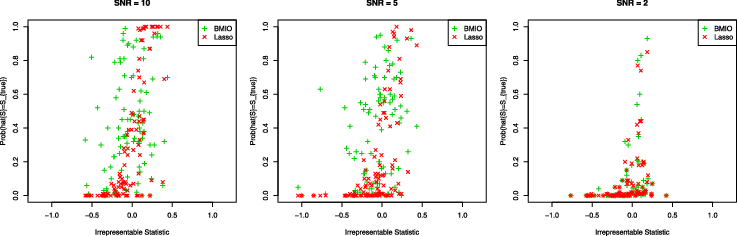}

\caption{Scatter plots for the sign probability $\mathbb{P}(\widehat
{S}=S_{0})$ against the Irrepresentable Statistic under different
signal-to-noise ratios.}
\label{simufigure2}
\end{figure}

Here we conduct a simulation study for the investigation. We generate
100 design matrices in which each row is i.i.d. from $\operatorname{MVN}(0,\Sigma
_{X})$, with $\Sigma_{X}\sim$ $\operatorname{Wishart}(I_{p\times p}, p,p)$ with
$p=30$. This setting is similar to the one used in Zhao and Yu's study.
Corresponding regression coefficients $\beta$ are generated in a way
that the first 5 entries of $\beta$ are i.i.d. from $\operatorname{Normal}(0,1)$, and
the rest of 25 entries are set to 0. Note that for some pairs $(X,\beta
)$, the Irrepresentable Condition (\ref{irr1}) will hold, but for some
pairs it will not hold. Zhao and Yu defined the irrepresentable
statistic by
\begin{equation}
\mathrm{Irr.stat} = 1 -
\Vert X_{S_{0}^{c}}^{T}X_{S_{0}}(X_{S_{0}}^{T}X_{S_{0}})^{-1}\operatorname
{sign}(\beta_{S_{0}})\Vert _{\infty}.
\end{equation}
The Irrepresentable Condition is considered to be violated if Irr.stat
is smaller than zero. We carry out 100 simulation runs and calculate
the irrepresentable statistic for each pair $(X,\beta)$. In each run,
we generate $n=100$ data points. We use $\sigma_{Y}^{2}= 0.05$ to
generate the error term $\varepsilon$, which is corresponding to SNR
$=10$. We then fit the regression model using the lasso estimation and
the BAVA-MIO estimation with these data points. For each pair $(X,\beta
)$, we calculate the model selection probability $P(\widehat
{S}_{0}=S_{0})$ based on counting the times of whether the estimated
sign vector matched the true sign vector throughout the whole
regularization paths.

We also carry out the same simulation experiment under SNR $= 5$, $2$
and~$1$. The scatter plots in Figure~\ref{simufigure2} show the
estimated model selection probability against Irr.stat under
signal-to-noise ratios SNR $ = 10$, $5$, and $2$. From these plots, we
can see that performances of the BAVA-MIO and the lasso estimations are
deteriorated when the signal-to-noise ratio is decreasing. However, we
also found in some circumstances the BAVA-MIO estimation can achieve
high model selection probabilities even when the Irrepresentable
Condition is violated, that is, Irr.stat is smaller than zero. In
Section~\ref{sec5}, we will provide a theoretical result to explain this
phenomenon. The second and fourth rows in Table~\ref{simutable10} show
the squared correlations between the estimated model selection
probability and the irrepresentable statistic. The squared correlations
for the BAVA-MIO estimation are relatively small in comparison with the
lasso estimation.

%
%t1 ###
\begin{table}
\caption{The sign probability $\mathbb{P}(\widehat{S}=S_{0})$ under
different signal-to-noise ratios. Each value is calculated by averaging
over 100 simulation runs, and the corresponding standard error~is~given
in the bracket. The term corr. in the second line of each panel is the
squared correlation between the sign probability and the
irrepresentable statistic. We~use~Kendall's $\tau$ for the correlation
calculation}\label{simutable10}
\begin{tabular*}{\textwidth}{@{\extracolsep{\fill}}lcccc@{}}
\hline
\multicolumn{1}{@{}l}{\textbf{Name}} & \multicolumn{1}{c}{$\mathbf{SNR} \bolds{= 10}$} & \multicolumn{1}{c}{$\mathbf{SNR} \bolds{= 5}$} & \multicolumn{1}{c}{$\mathbf{SNR} \bolds{= 2}$} &
\multicolumn{1}{c@{}}{$ \mathbf{SNR} \bolds{= 1}$} \\
\hline
BMIO & 0.398 (0.030) & 0.338 (0.030) & 0.077 (0.018) & 0.023 (0.006)\\
corr. & 0.052\phantom{ (0.030)} & 0.047\phantom{ (0.030)} &0.180\phantom{ (0.030)} & 0.110\phantom{ (0.030)}\\[3pt]
Lasso & 0.314 (0.047) &0.203 (0.030) &0.072 (0.016) &0.022 (0.006)\\
corr. &0.418\phantom{ (0.030)} &0.232\phantom{ (0.030)} &0.194\phantom{ (0.030)} & 0.107\phantom{ (0.030)}\\
\hline
\end{tabular*}
\end{table}

%s5 ###
\section{Asymptotic analysis}\label{sec5}
In this section, we will derive asymptotic results for the BAVA-MIO
estimator. When deriving the asymptotic results, we will consider a
situation in which the number of parameters $p$ is an increasing
function of the number of samples $n$. For practical purposes, we will
focus on the case $p=p(n)\propto n^{\alpha}$, where $\alpha> 0$. The
first asymptotic result gives a theoretical explanation for the
invariance of the BAVA-MIO estimator under the Irrepresentable
Condition. The second result is on the posterior model consistency
related to the hierarchical Bayesian formulation (\ref{2}), and the
third result shows the estimation consistency of the BAVA-MIO estimator.

%s5.1 ###
\subsection{Sign consistency}\label{sec51}
Before stating the result of sign consistency, we give some notation
definitions first. We use the same definitions given in Section~\ref{sec42} for
$S_{0}$, $S_{0}^{c}$, $\beta_{S_{0}}$, $\beta_{S_{0}^{c}}$, $X_{S_{0}}$
and $X_{S_{0}^{c}}$. Further let $\mathcal{S}$ denote the space that
$S_{0}$ belongs to. For a symmetric matrix $C$, let $\Lambda_{\min}(C)$
and $\Lambda_{\max}(C)$ denote the smallest and the largest
eigenvalues, respectively.

Our result on the sign consistency of the BAVA-MIO estimator is based
on the following simplification: the variable $\sigma^{2}$ is fixed and
the term $\rho_{\lambda,\kappa,\sigma^{2}}$ in (\ref{6}) is treated as
a constant. For simplicity, we let $\rho=\rho_{\lambda,\kappa,\sigma
^{2}}$. Now define
%
%e22 ###
\begin{equation}
\label{sparsist01}
\widehat{\beta}^{\tau_{3}} = \arg\min_{\beta}\Vert y-X\beta\Vert _{2}^{2} +
\lambda\Vert \beta\Vert _{2}^{2} + \rho\sum_{j=1}^{p}\frac{\log(1+\tau
_{3}^{-1}|\beta_{j}|)}{\log(1+\tau_{3}^{-1})}.
\end{equation}
Note that the log-sum function on the right-hand side of (\ref{sparsist01}) becomes $\Vert \beta\Vert _{0}$ if $\tau_{3}\rightarrow0$, and
$\widehat{\beta}^{\tau_{3}}$ in this sense can be seen as the BAVA-MIO
estimator. Now define
%
%e23 ###
\begin{equation}
\label{sparsist03}
E_{0,\tau_{3}} = \{\beta\dvtx  \operatorname{sign}(\beta_{j})=\operatorname
{sign}(\widehat{\beta}_{j}^{\tau_{3}}) \mbox{ for }  j=1,2,\ldots
,p\},
\end{equation}
that is, the event of sign consistency for the estimator $\widehat{\beta
}^{\tau_{3}}$ in (\ref{sparsist01}). For practical purposes, further
define $C_{SS_{0}}=n^{-1}(X_{S_{0}}^{T}X_{S_{0}}+\lambda I)$,
$C_{S^{c}S_{0}}=n^{-1}X_{S_{0}^{c}}^{T}X_{S_{0}}$,
$D_{S_{0}}=n^{-1/2}X_{S_{0}}^{T}\varepsilon$ and
$D_{S_{0}^{c}}=n^{-1/2}X_{S_{0}^{c}}^{T}\varepsilon$. In the following, we
give some assumptions that will be used in deriving the asymptotic results.
\begin{ass}\label{ass1}
For $C_{SS}=n^{-1}(X_{S}^{T}X_{S}+\lambda
I)$ and any $S\in\mathcal{S}$, the maximum eigenvalue $\Lambda_{\max
}(C_{SS})$ and the minimum eigenvalue $\Lambda_{\min}(C_{SS})$ satisfy
the following condition:
\[
0\leq c_{1}<\Lambda_{\min}(C_{SS})\leq\Lambda_{\max}(C_{SS})\leq
c_{2}<\infty.
\]
\end{ass}
\begin{ass}\label{ass2} For the vector $X^{T}\varepsilon$,
$\Vert X^{T}\varepsilon\Vert _{1}=O(p)$.
\end{ass}
\begin{ass}\label{ass3} For parameter $\lambda$, we assume $0\leq
\lambda<\infty$. For parameter $\rho$, we assume $0\leq\rho$ and $\rho
n^{-1/2}\rightarrow0$.
\end{ass}

Assumption~\ref{ass1} is a special case of the Restricted Eigenvalue Assumption
stated in Bickel et al.~\cite{bickeletal09}. It implies that the
inverse of $C_{SS_{0}}$ exists and the ratio $[\Lambda_{\max
}(X_{S}^{T}X_{S})+\lambda]/[\Lambda_{\min}(X_{S}^{T}X_{S})+\lambda]\leq
c_{2}/c_{1}$ is bounded from above for any $S\in\mathcal{S}$.
Assumption~\ref{ass2} is equivalent to the statement that
$n^{-1/2}\Vert X^{T}\varepsilon\Vert _{1}$ is bounded from some quantity
proportional to $pn^{-1/2}$ as $n\rightarrow\infty$, which further
implies $\Vert D_{S_{0}}\Vert _{1}$ and $\Vert D_{S_{0}^{c}}\Vert _{1}$ are bounded
from the quantity as well.
\begin{theorem}
\label{theorem1}
Given that Assumptions~\ref{ass1} to~\ref{ass3} hold, if the number of covariates
$p\propto n^{\alpha}$, $0<\alpha< 1/2$, and $\tau_{3}\propto n^{-1}$,
then we have
\[
\mathbb{P}(E_{0,\tau_{3}})\rightarrow1
\]
as $n\rightarrow\infty$.
\end{theorem}

The proof is given in Appendix C of Supplementary Material~\cite{sup}. The proof
will start by exploring the KKT conditions associated to the
minimization problem stated in~(\ref{sparsist01}). Note that in Theorem
\ref{theorem1} we do not assume that the Irrepresentable Condition
should hold. Indeed, as stated in Corollary C1 in Appendix C, even if
the Irrepresentable Condition is violated, Theorem~\ref{theorem1} will
still hold given that some mild condition is imposed.

%s5.2 ###
\subsection{Posterior model consistency}\label{sec52}
We give notation definitions first. Let $y^{n}=(y_{1},y_{2},\ldots
,y_{n})$. The notation $y^{n}$ emphasizes the fact that the number of
entries in the observed response vector $y$ is $n$. Further let
$\mathcal{M}_{S}$ denote the model characterized by the index set $S$.
Under a Bayesian framework, $\mathcal{M}_{S}$ usually refers to the
sampling density, and posterior model consistency is defined as $\mathbb
{P}(\mathcal{M}_{S_{0}}| y^{n})\rightarrow1$ as $n\rightarrow\infty$,
where $\mathcal{M}_{S_{0}}$ can be seen as the ``true model,'' or the
true sampling density that a sample comes from. The posterior model
consistency states that the posterior probability will put all its mass
on $\mathcal{M}_{S_{0}}$ as the sample size goes to infinity. Note that
in general, multiple true models are allowed under Bayesian frameworks,
that is, $S_{0}$ may not be unique. However, for simplicity we only pay
attention on the situation in which there is only one true model.

Note that the posterior probability $\mathbb{P}(\mathcal{M}_{S^{\prime
}}| y^{n})$ with $S^{\prime}\in\mathcal{S}$ can be expressed in terms
of Bayes factors by
%
%e24 ###
\begin{eqnarray}
\label{consist1}
\mathbb{P}(\mathcal{M}_{S^{\prime}}| y^{n})&=&\frac{f(y^{n}| \mathcal
{M}_{S^{\prime}})f(\mathcal{M}_{S^{\prime}})}{\sum_{S\in\mathcal
{S}}f(y^{n}| \mathcal{M}_{S})f(\mathcal{M}_{S})}
\nonumber
\\[-8pt]
\\[-8pt]
\nonumber
&=&\frac{\mathrm
{BF}(\mathcal{M}_{S^{\prime}},\mathcal{M}_{S_{0}};y^{n})f(\mathcal
{M}_{S^{\prime}})}{\sum_{S\in\mathcal{S}}\mathrm{BF}(\mathcal
{M}_{S},\mathcal{M}_{S_{0}};y^{n})f(\mathcal{M}_{S})}.
\end{eqnarray}
The formulation (\ref{consist1}) implies that the event $\mathbb
{P}(\mathcal{M}_{S_{0}}| y^{n})\rightarrow1$ is equivalent to the
events $\mathrm{BF}(\mathcal{M}_{S},\mathcal{M}_{S_{0}})\rightarrow0$
for all $S\in\mathcal{S}$ and $S\neq S_{0}$, given that the probability
$f(\mathcal{M}_{S_{0}})$ is bounded from zero. It turns out that to
examine whether the posterior probability is consistent at the true
model $\mathcal{M}_{S_{0}}$ is the same as to examine whether Bayes
factors between other models and the true model will approach to zero
or not.

Here, we make some assumptions on the Bayesian formulation (\ref{2})
before stating the main result of posterior model consistency.
\begin{ass}\label{ass4} The prior probability on the true model
$\mathcal{M}_{S_{0}}$ is bounded away from zero, that is, $f(\mathcal
{M}_{S_{0}})>0$.
\end{ass}
\begin{ass}\label{ass5}
We assume $\lambda^{-1}\tau_{2}<\infty$.
\end{ass}
\begin{ass}\label{ass6} The condition
\begin{equation}
\frac{(y^{n})^{T}y^{n}}{\Lambda_{\min}(X_{S_{0}}X_{S_{0}}^{T})}<
(y^{n})^{T}(X_{S}X_{S}^{T}+\lambda I)^{-1}y^{n}\nonumber
\end{equation}
holds for all $n\in\mathbb{N}^{+}$, $S\in\mathcal{S}\setminus S_{0}$
and $0\leq\lambda<\infty$.
\end{ass}

Assumption~\ref{ass4} states that the true model should always have positive
mass under the prior. It is a reasonable assumption since otherwise by
Bayes' theorem the posterior probability of $\mathcal{M}_{S_{0}}$ will
be zero. In addition, Assumption~\ref{ass6} is a technical condition which
ensures that the ratio $(y^{n})^{T}(X_{S_{0}}X_{S_{0}}^{T}+\lambda
I)^{-1}y^{n}/[(y^{n})^{T}(X_{S}X_{S}^{T}+\lambda I)^{-1}y^{n}]$ is
smaller than 1. The assumption will be useful in proving the
convergence of the Bayes factor $\operatorname{BF}(\mathcal{M}_{S},\mathcal{M}_{S_{0}};y^{n})$.
\begin{theorem}
\label{theorem2}
Given Assumptions~\ref{ass4} to~\ref{ass6} hold and the number of covariates $p\propto
n^{\alpha}$, the inequality
\[
\mathbb{P}(\mathcal{M}_{S_{0}}| y^{n})\geq1 - c_{3}\exp\biggl\{-\frac
{n^{\alpha}}{2}(n^{1-\alpha}\xi-c_{11}\log4)\biggr\}
\]
will hold for some constants $0\leq c_{3} < \infty$, $\xi> 0$,
$0<c_{11}<\infty$ and $n^{*} > 0$ for all $n>n^{*}$. Therefore, for
$0<\alpha< 1$, $\mathbb{P}(\mathcal{M}_{S_{0}}| y^{n})\rightarrow1$
as $n\rightarrow\infty$.
\end{theorem}

The proof of Theorem~\ref{theorem2} is given in Appendix D of
Supplementary Material~\cite{sup}.

%s5.3 ###
\subsection{Estimation consistency}\label{sec53}
Using (\ref{sparsist01}), the BAVA-MIO estimator can be defined as
$\widehat{\beta}_{\mathrm{BMIO}}=\lim_{\tau_{3}\rightarrow0}\widehat{\beta
}^{\tau_{3}}$. Now we deal with the estimation consistency of $\widehat
{\beta}_{\mathrm{BMIO}}$ under a frequentist's framework. We will derive
an asymptotic bound for the $l_{2}$ distance between $\widehat{\beta
}_{\mathrm{BMIO}}$ and $\beta_{0}$ and show that the $l_{2}$ distance
converges to~0 as $n\rightarrow\infty$. Let $\beta_{0}$ denote the
coefficient vector corresponding to\vadjust{\goodbreak} the true model $\mathcal
{M}_{S_{0}}$. Define the expected $l_{2}$ distance between estimator
$\widehat{\beta}$ and the true coefficient~$\beta_{0}$ by
\[
\mathbb{E}_{Y}[\Vert \widehat{\beta}-\beta_{0}\Vert _{2}^{2}] = \int\Vert \widehat
{\beta}-\beta_{0}\Vert _{2}^{2}f(y| \beta_{0})\,dy,
\]
where $f(y| \beta_{0})$ is the sampling density parametrized by the
true parameter $\beta_{0}$. Before deriving the asymptotic result, we
make some finite moment assumptions on the true parameters $(\beta
_{0},\sigma^{2})$.
\begin{ass}\label{ass7} There exist finite constants $c_{4}>0$
and $c_{5}>0$ such that $\beta_{0,j}^{2}<c_{4}$ for $j=1,2,\ldots,p$,
and $\sigma^{2}<c_{5}$.
\end{ass}

Assumption~\ref{ass7} ensures that parameter $\beta_{0}$ and parameter $\sigma
^{2}$ are bounded away from above as the sample size $n$ goes large.
\begin{theorem}
\label{theorem3}
Given Assumptions~\ref{ass1},~\ref{ass3} and~\ref{ass7} hold and the number of covariates
$p\propto n^{\alpha}$ with $\alpha> 0$, the inequality
\begin{equation}
\mathbb{P}(\Vert \widehat{\beta}_{\mathrm{BMIO}}-\beta
_{0}\Vert _{2}^{2} > \xi_{n})\leq c_{13}\exp\{-\log(n^{1-\alpha
}\xi_{n})\}
\end{equation}
will hold for some positive finite constant $c_{13}$ and $\xi_{n}$.
Assume $\xi_{n}\geq0$ is decreasing with $n$, that is, $\xi
_{n}\rightarrow0$ as $n\rightarrow\infty$. Let $\xi_{n} \propto
n^{-\alpha^{*}}$ for some $\alpha^{*}>0$. Then with the condition
$0<\alpha^{*}<\alpha<1/2$, $\mathbb{P}(\Vert \widehat{\beta}_{\mathrm{BMIO}}-\beta_{0}\Vert _{2}^{2} > \xi_{n})\rightarrow0$ as $n\rightarrow
\infty$.
\end{theorem}

The proof of Theorem~\ref{theorem3} is given in Appendix E of
Supplementary Material~\cite{sup}.

%s6 ###
\section{An extension to generalized linear models}\label{sec6}
Here we extend the proposed method, the BAVA-MIO, to parameter
estimation in the generalized linear models. Consider the density of
the exponential family
%
%e25 ###
\begin{equation}
\label{glm0}
f(y| \theta,\varphi)=\exp\biggl\{\frac{y\theta-b(\theta)}{\varphi
}+d(y,\varphi)\biggr\},
\end{equation}
where $\theta$ is a parameter characterizing mean of the distribution
and $\varphi$ is a parameter characterizing dispersion of the
distribution. Under the exponential family (\ref{glm0}), variable $Y$
has properties such that $\mathbb{E}(Y| \theta,\varphi)=b^{\prime
}(\theta)$, $\operatorname{Var}(Y| \theta,\varphi)=b^{\prime\prime}(\theta)\varphi$.
Now let $\nu=b^{\prime}(\theta)$. For a generalized linear model, there
exists a link function~$\eta$ such that $\eta(\nu)=x^{T}\beta$. The
link function gives a flexible connection between the mean $\nu$ and
the predictor $x^{T}\beta$, and a valid regression can be formulated
under this parametrization. In addition, $\nu$ is parametrized by
$\theta$, therefore by inverse mapping, we can express $\theta$ as a
function of $x^{T}\beta$. We write $ \theta= \theta(x^{T}\beta)$.

For a practical inference concern, we will not assign a prior on
$\varphi$ in the following Bayesian hierarchical formulation. We only
assign priors on regression coefficients~$\beta$ and covariate indices\vadjust{\goodbreak}
$\gamma$. The inference concern arises from the fact that the
estimation of $\varphi$ is dependent on the function $d(y,\varphi)$,
and in general, $d(y,\varphi)$ is case dependent. We will launch an
investigation on how to assign a prior on $\varphi$ in the future, but
at present we only focus on inference based on priors on $\beta$ and
$\gamma$. Now consider the logarithm of the joint density function
%
%e26 ###
\begin{eqnarray}
\label{glm1}
-\log f(\beta,\gamma| X,y,\varphi,\lambda,\kappa)&=&-\sum
_{i=1}^{n}\biggl\{\frac{y_{i}\theta_{i}(x_{i}^{T}\beta)-b[\theta
_{i}(x_{i}^{T}\beta)]}{\varphi}+d(y_{i},\varphi)\biggr\}\nonumber\\
& &{}+\frac{\lambda}{2\varphi}\sum_{j=1}^{p}\gamma_{j}\beta
_{j}^{2}\\
& &{}+\frac{1}{2}\sum_{j=1}^{p}\gamma_{j}\log\biggl\{\frac{2\pi\varphi
(1-\kappa)^{2}}{\lambda\kappa^{2}}\biggr\} + \mathrm{const}.\nonumber
\end{eqnarray}
The first term in (\ref{glm1}) is the logarithm of joint sampling
density over $i=1,2,\ldots,n$, and the second and third terms are
logarithms of the priors on $\beta$ and covariate indices $\gamma$,
respectively. To modify (\ref{glm1}) for BAVA-MIO estimation, we first
multiply (\ref{glm1}) with $\varphi$. We then apply a
majorization--minimization technique to obtain an approximation to the
$l_{0}$ norm penalty. The BAVA-MIO estimator of $\beta$ is defined as
the minimizer of the approximate objective, which can be obtained by
the following iteration scheme:
%
%e27 ###
\begin{eqnarray}
\label{glm11}
\widehat{\beta}^{(m+1)}&=&\arg\min\Biggl\{-\sum_{i=1}^{n}\{y_{i}\theta
_{i}(x_{i}^{T}\beta)-b[\theta_{i}(x_{i}^{T}\beta)]\}
\nonumber
\\[-8pt]
\\[-8pt]
\nonumber
&&\hspace*{71pt}{}+ \frac{\lambda
}{2}\Vert \beta\Vert _{2}^{2} + \rho\Vert \widehat{\phi}^{(m)}\beta\Vert _{1}\Biggr\}
,
\end{eqnarray}
where $\rho= \varphi[\log2\pi\varphi(1-\kappa)^{2}(\lambda\kappa
^{2})^{-1}]/2$ and $\widehat{\phi}^{(m)}=\lim_{\tau_{3}\rightarrow
0}[\log(1+\tau_{3}^{-1})\times (|\widehat{\beta}^{(m)}|+\tau_{3})]^{-1}$.
Further, by differentiating (\ref{glm11}) with respect to $\beta$,
and setting the derivatives to zero, we obtain the subgradient
equations of $\beta$, which are given by
%
%e28 ###
\begin{equation}
\label{glm2}
-X^{T}Wr + \lambda\beta+ g_{\beta}\rho\widehat{\phi}^{(m)} = 0,
\end{equation}
where $r=(y - \nu)\eta^{\prime}(\nu)$, $W = \operatorname{diag}\{[\eta^{\prime}(\nu
_{1})^{2}]b^{\prime\prime}(\theta_{1}),\ldots,[\eta^{\prime}(\nu
_{n})^{2}]b^{\prime\prime}(\theta_{n})\}^{-1}$, $\nu=(\nu_{1},\nu
_{2},\ldots,\nu_{n})$ with $\nu_{i}=b^{\prime}(\theta_{i})$ and
$g_{\beta}=(g_{\beta_{1}},g_{\beta_{2}},\ldots,g_{\beta_{p}})$ is the
subgradient vector of $\Vert \beta\Vert _{1}$ such that $g_{\beta_{j}}=1$ if
$\beta_{j}> 0$, $g_{\beta_{j}}=-1$ if $\beta_{j}<0$ and $g_{\beta
_{j}}\in[-1,1]$ if $\beta_{j}=0$. The term $X^{T}Wr$ in (\ref{glm2})
is a standard result in parameter estimation of the generalized linear
models, and its derivation can be found in~\cite{mccullaghandnelder89}.
The term $X^{T}Wr$ allows us to formulate an iteration scheme to
approximate the solution of the subgradient equations (\ref{glm2}).
Here we will use the iteration scheme
%
%e29 ###
\begin{equation}
\label{glm3}
\qquad(\widehat{\beta}^{*})^{(m+1)}=\arg\min_{\beta}\biggl\{\frac
{1}{2}\bigl\Vert U^{(m)}\bigl(z^{(m)} - X\beta\bigr)\bigr\Vert _{2}^{2} + \frac{\lambda}{2}\Vert \beta
\Vert _{2}^{2} + \rho\bigl\Vert\widehat{\phi}^{(m)}\beta\bigr\Vert_{1}\biggr\},
\end{equation}
where
\begin{eqnarray*}
\label{glm4}
z^{(m)}&=&r^{(m)} + \eta^{(m)},\\
U^{(m)}&=&(W^{1/2})^{(m)},
\end{eqnarray*}
to approximates the solution of the subgradient equations (\ref{glm2}). The $j$th element of the iteration scheme (\ref{glm3}) can
be obtained by further carrying out the following soft-thresholding
scheme coordinatewise:
%
%e30 ###
\begin{equation}
\label{glm5}
\quad(\tilde{\beta}_{j}^{*})^{(m+1,l+1)}=\Biggl(\sum
_{i=1}^{n}w_{ii}^{(m)}x_{ij}^{2} + \lambda\Biggr)^{-1}\operatorname{ST}\Biggl(\sum
_{i=1}^{n}x_{ij}w_{ii}^{(m)}\tilde{v}_{i,-j}^{(m,l)},\rho\tilde{\phi
}_{j}^{(m)}\Biggr),
\end{equation}
where $w_{ii}^{(m)}$ is the $i$th diagonal term of $W^{(m)}$, $\tilde
{v}_{i,-j}^{(m,l)}=z_{i}^{(m)}-\sum_{j^{\prime}\neq j}x_{ij}\tilde{\beta
}_{j^{\prime}}^{*}$ with $\tilde{\beta}_{j^{\prime}}^{*}=(\tilde{\beta
}_{j^{\prime}}^{*})^{(m+1,l+1)}$ for $j^{\prime}=1,2,\ldots,j-1$ and
$\tilde{\beta}_{j^{\prime}}^{*}=(\tilde{\beta}_{j^{\prime
}}^{*})^{(m,l)}$ for $j^{\prime}=j+1,j+2,\ldots,p$, and $\operatorname{ST}(a,b)$ is
the soft-thresholding operator defined by $\operatorname{ST}(a,b)=\operatorname{ sign}(a)(|a|-b)_{+}$.

We now conduct a simulation study to assess the performance of the
BAVA-MIO estimation. We take logistic regression as the example. For
the true model, we assume $Y_{i}\sim$ $\operatorname{Bernoulli}(\zeta_{i})$, where
$\zeta_{i}$ is parametrized in terms of predictor $x_{i}^{T}\beta$ via
the link function $\log[(\zeta_{i})/(1-\zeta_{i})]$. We further let the
number of covariates $p=120$. For the regression coefficients $\beta$,
we generate the $j$th entry $\beta_{j}$ from $\operatorname{Normal}(0,1)$ for
$j=1,2,\ldots,10$, and let the rest of 110 $\beta_{j}'s$ equal to zero.
We simulate covariate vector $x_{i}$ i.i.d. from $\operatorname{MVN}(0,\Sigma_{X})$.
We consider three $\Sigma_{X}$'s, the same as those described in
Section~\ref{sec41}, to generate the covariate vectors. With $\beta$ and
$x_{i}$, we simulate $Y_{i}$ from $\operatorname{Bernoulli}(\zeta_{i})$ for the cases
of $n=100$ and $n=200$. With three different values for $\Sigma_{X}$
and two different values for $n$, we have six scenarios in the
simulation study. For each scenario, we generate 100 simulation runs.
In each simulation run, we apply the BAVA-MIO estimation to fit a
logistic regression model. We let hyperparameter $\lambda=n^{-1/2}$ for
all estimations. Note that for a Bernoulli variable, a closed form
representation for the Bayes factor does not exist, therefore we only
use ten fold cross validation for tuning parameter selection. For
comparison purposes, we also carry out the lasso estimation using R~package ``glmnet'' and use ten fold cross validation for tuning
parameter selection. We collect four performance measures, the same as
those described in Section~\ref{sec41}, at each simulation run. Average values
of the four performance measures over the 100 simulation runs are given
in Table~\ref{glmtable1}. From these tables we can see that the
BAVA-MIO estimation in general has slightly larger values in PMSE than
the lasso estimation has, but it gives far fewer number of selected
covariates, more accurate results in covariate selection, and in some
circumstances, better parameter estimation than the lasso estimation.
%
%t2 ###
\begin{table}
\caption{Results of BAVA-MIO GLM estimation. Each value is calculated
by averaging over 100 simulation runs, and the corresponding standard
error is given in the bracket. BMIO-CV: the BAVA-MIO estimation using
ten-fold cross validation; lasso: the lasso estimation. The top panel:
covariance matrix with off-diagonal terms equal to $0$; The middle
panel: covariance matrix with off-diagonal terms equal to $0.5$; The
bottom panel: covariance matrix with off-diagonal terms following a
specified covariance structure}\label{glmtable1}
\begin{tabular*}{\textwidth}{@{\extracolsep{\fill}}lccccc@{}}
\hline
& \multicolumn{1}{c}{$\bolds{n}$}&
\multicolumn{1}{c}{\textbf{PMSE}} &
\multicolumn{1}{c}{$\bolds{l}_{\bolds{2}}$\textbf{-dis}} &
\multicolumn{1}{c}{\textbf{S-FPR}} &
\multicolumn{1}{c@{}}{$\bolds{|\widehat{S}|}$}\\
\hline
BMIO-CV &100 &0.186 (0.004) & 0.039 (0.002) & 0.305 (0.031)& 11.39 (1.788)\\
GLM-lasso&100 &0.173 (0.003) & 0.044 (0.004)& 0.680 (0.012)& 21.34 (1.050)\\
BMIO-CV&200 & 0.145 (0.003)& 0.018 (0.001) & 0.176 (0.021) & \phantom{0}7.54 (0.549)\\
GLM-lasso&200 & 0.144 (0.002)& 0.026 (0.001) & 0.694 (0.011) & 27.65
(1.141)\\[3pt]
BMIO-CV &100 &0.180 (0.004)& 0.055 (0.003)& 0.455 (0.031)& 13.86 (1.841)\\
GLM-lasso&100 & 0.169 (0.003)& 0.052 (0.003)& 0.654 (0.018)& 17.54 (0.966)\\
BMIO-CV&200 & 0.157 (0.003) &0.027 (0.002) & 0.327 (0.025)& \phantom{0}8.58 (0.564)\\
GLM-lasso&200 & 0.154 (0.003) & 0.030 (0.001) & 0.654 (0.012) & 20.67
(0.908)\\[3pt]
BMIO-CV&100 &0.180 (0.004) & 0.046 (0.003) & 0.271 (0.028)& \phantom{0}8.28 (1.187)\\
GLM-lasso&100 & 0.170 (0.003)& 0.047 (0.003)& 0.668 (0.016)& 19.18 (0.914)\\
BMIO-CV&200 & 0.152 (0.004) & 0.022 (0.001) & 0.203 (0.023)&\phantom{0}7.19 (0.478)\\
GLM-lasso&200 & 0.150 (0.003) & 0.027 (0.001) & 0.675 (0.014)& 23.50 (1.045)\\
\hline
\end{tabular*}
\end{table}

%s7 ###
\section{Real data examples}\label{sec7}
In this section, we present two real data analyses. We will apply
methods developed in Section~\ref{sec3} and Section~\ref{sec6} to estimate parameters in
regression models.

%s7.1 ###
\subsection{Diabetes data}\label{sec71}
The Diabetes data contains a measure on disease progression and 10
covariates: age, sex, the BMI index, blood pressure and six related
variables for 442 diabetes patients. In our analysis, each covariate
has been rescaled to have mean zero and variance 1, and the response
variable has been centered around its mean. All estimations are based
on the rescaled covariates and centered response variable. For hyperparameters,
we let $(\tau_{1},\tau_{2})=(1,1)$ and $\lambda=0.2\times
\sqrt{p\log(p)/n}\approx0.049$. We perform two BAVA-MIO estimations.
The first one uses the Bayes factor (BMIO-BF) while the second one uses
ten fold cross validation (BMIO-CV) for tuning parameter selection. The
results are shown in the first two columns of Table~\ref{realtable1}.
From the results, we can see the BMIO-BF estimation leads to a
covariate selection sparser than its counterpart using ten fold cross
validation. We also run another 100 estimations based on sampling half
of the 442 subjects\vadjust{\goodbreak} without replacement to calculate the inclusion
probabilities for the 10 covariates. For each covariate, the inclusion
probability is defined as the proportion of occurrences of nonzero
estimated values appearing in the 100 subsampling estimations. We
compare the results from the BAVA-MIO estimations with the results from
three other estimation approaches: g-prior, hyper-g and BIC. All the
three estimations are carried out using R package ``BAS,'' which is
developed by Clyde, Ghosh and Littman~\cite{clydeetal09} as the
companion software to the paper of Liang et al.~\cite{liangetal07}.
These results are shown in the last three columns of Table~\ref{realtable1}. For the three estimations using the BAS package, we
report the models estimated with the highest marginalized likelihood.
The results show that the estimation based on BAVA-MIO using the Bayes
factor has relative sparse covariate selection among the five proposed
approaches. Among the 10 inclusion probabilities estimated via the
BMIO-BF estimation, only four are above 0.5, compared to five for the
BMIO-CV estimation, six for the g-prior and the BIC estimations, and
seven for the hyper-g estimation.
%
%t3 ###
\begin{table}
\caption{Estimation results based on the Diabetes data. The value in
the bracket\break is the inclusion probability of the covariate based on the
100\break subsampling estimations. For g-prior, hyper-g and BIC, the value\break in
the bracket is the posterior inclusion probability of the
covariate}\label{realtable1}
\begin{tabular*}{\textwidth}{@{\extracolsep{\fill}}ld{3.8}d{3.8}d{3.8}d{3.8}d{3.8}@{\hspace*{-3pt}}}
\hline
\multicolumn{1}{@{}l}{\textbf{Name}} & \multicolumn{1}{c}{\textbf{BMIO-BF}\hspace*{5pt}} &\multicolumn{1}{c}{\textbf{BMIO-CV}\hspace*{5pt}} &
\multicolumn{1}{c}{\textbf{g-prior}\hspace*{6pt}} &
\multicolumn{1}{c}{\textbf{hyper-g}\hspace*{6pt}} & \multicolumn{1}{c@{}}{\textbf{BIC}\hspace*{2pt}}\\
\hline
age & 0.00\ (0.01) & 0.00\ (0.30) & 0.00\ (0.11) & 0.00\ (0.33) & 0.00
\ (0.05)\\
sex & -11.23\ (0.49) & -11.08\ (0.64) &-10.64\ (0.99) & -8.02\ (0.97)
&-10.71\ (0.98)\\
bmi & 24.92\ (1.00)& 25.06\ (1.00)& 24.96\ (1.00)& 19.00\ (1.00)& 25.37
\ (1.00)\\
map & 15.54\ (0.86) & 15.01\ (0.92) & 15.29\ (1.00) & 11.55\ (1.00) & 15.53
\ (1.00)\\
tc & 0.00\ (0.03) & 0.00\ (0.33) &-16.62\ (0.71) &-13.54\ (0.75) & 0.00
\ (0.57)\\
ldl & 0.00\ (0.10) & 0.00\ (0.28) & 8.51\ (0.50) & 6.51\ (0.59) & 0.00
\ (0.38)\\
hdl & -13.76\ (0.69) & -11.20\ (0.81) & 0.00\ (0.49) & 0.00\ (0.57) & -7.29
\ (0.57)\\
tch & 0.00\ (0.01) & 0.00\ (0.27) & 0.00\ (0.30) & 0.00\ (0.48) & 0.00
\ (0.20)\\
ltg & 22.59\ (1.00)& 25.72\ (1.00)& 29.13\ (1.00)& 22.29\ (1.00)& 28.31
\ (1.00)\\
glu & 0.00\ (0.10) & 3.44\ (0.47) & 0.00\ (0.17) & 0.00\ (0.41) & 0.00
\ (0.07)\\
\hline
\end{tabular*}
\end{table}

%s7.2 ###
\subsection{Golub's Leukemia data}\label{sec72}
The Leukemia gene expression data, adopted from R package
``golubEsets,'' is originally from~\cite{golubetal99}. It consists of
gene expression profiles for 72 Leukemia patients, of which 47 are
diagnosed with acute lymphoblastic leukemia (ALL) and 25 are diagnosed
with acute myeloid leukemia (AML). Each profile has 7,129 gene
expression values measured by Affymetrix Hgu6800 chips. The data set is
further divided into the training set, which consists of 27 ALL
patients and 11 AML patients, and the test set, which consists of 20
ALL patients and 14 AML patients. Our aim is to identify a patient's
disease type with a small set of genes. The data set is processed as
follows. The disease type is labeled with 0 for the acute lymphoblastic
leukemia and 1 for the acute myeloid leukemia. Each covariate is first
rescaled to have a range greater than or equal to zero. Then it is
under a suitable logarithm transform before rescaled again to have mean
0 and variance 1. For the classification rule construction, we apply
the BAVA-MIO estimation to fit logistic regression models with the
training data. We parametrize hyperparameter $\lambda=\lambda^{*}\sqrt
{p\log p/n}$ and perform three estimations with $\lambda^{*}=0.05,0.1$
and $0.5$. The tuning parameter is selected via five fold cross
validation and the resulting estimates are termed BMIO-CV I, BMIO-CV II
and BMIO-CV III, respectively. With estimated regression coefficients,
we calculate the label probability for each patient, and classifying
those with label probabilities smaller than 0.5 to the acute
lymphoblastic leukemia group, and those with label probabilities
greater than 0.5 to the acute myeloid leukemia group. The corresponding
classification results are reported in Table~\ref{realtable2}, along
with classification results on the same data set done by Golub et al.
\cite{golubetal99} and four other estimation approaches~\cite{zouandhastie05,parkandhastie07,fanandfan08,fanandlv08} aiming to
tackle high-dimensional classification problems. The results show that
BAVA-MIO-based classification rules tend to use less numbers of genes
in identifying a patient's disease type. However, even with smaller
numbers of genes, the BAVA-MIO-based classification rules can still
generate results that are comparable with those provided by other
benchmark methods.
%
%t4 ###
\begin{table}
\caption{Classification results for Golub's gene expression data}\label{realtable2}
\begin{tabular*}{\textwidth}{@{\extracolsep{\fill}}lccc@{}}
\hline
\multicolumn{1}{@{}l}{\textbf{Method}} &
\multicolumn{1}{c}{\textbf{CV-error}} &
\multicolumn{1}{c}{\textbf{Test-error}} &
\multicolumn{1}{c@{}}{\textbf{\# of genes}}\\
\hline
Golub et al.~\cite{golubetal99} & 3$/$38& 4$/$34 & 50\\
Elastic Net (Zou and Hastie~\cite{zouandhastie05})& 3$/$38&0$/$34 &45\\
$l_{1}$-pen GLM (Park and Hastie~\cite{parkandhastie07})& 1$/$38& 2$/$34& 23\\
SIS-SCAD-LD (Fan and Lv~\cite{fanandlv08}) & 0$/$38& 1$/$34&16\\
FAIR (Fan and Fan~\cite{fanandfan08}) &1$/$38 &1$/$34 & 11\\
BMIO-CV I & 1$/$38 & 1$/$34&\phantom{0}8\\
BMIO-CV II &1$/$38& 1$/$34&\phantom{0}9\\
BMIO-CV III &1$/$38&0$/$34&23\\
\hline
\end{tabular*}
\end{table}

%s8 ###
\section{Concluding remarks}\label{sec8}

One important issue to which we did not pay much attention is the impacts of
hyperparameters on estimation results. Here we provide some possible
modifications in addressing this issue. First, an equally spaced grid
may be constructed for hyperparameter $\lambda$ so that the estimation
procedure can be carried out along the grids on $\lambda$ and $\Psi
(\kappa)$. Another possible modification is to drop the prior
assumption on $\sigma^{2}$ and treat it as a constant. In this approach
the impact of $\sigma^{2}$ on parameter estimation can be dealed
together with the tuning parameter~$\Psi(\kappa)$. This approach has
been adopted in Section~\ref{sec6} for parameter estimation in the generalized
linear models.\vadjust{\goodbreak}

\section*{Acknowledgments}

We thank the Associate Editor and the reviewer for their invaluable
comments. We are grateful to Dr. Chun-houh Chen for his encouragement
and helpful suggestions. We also thank Professor Yuan-chin Chang, Dr.
Ting-Li Chen, Dr. Kai-Ming Chang and Mr. Yu-Min Yen for helpful
comments.

\begin{supplement} %[id=suppA]
\stitle{Supplement File}
\slink[doi]{10.1214/11-AOS884SUPP} %[doi,text={...}] - jei reikia
%suskaldyti doi
\sdatatype{.pdf}
\sfilename{aos884\_suppl.pdf}
\sdescription{In Supplementary Material, we provide brief discussions
on the log-sum function, connections with other approaches, derivation
of the soft-thresolding operator, and proofs of Theorems~\ref{theorem1},~\ref{theorem2} and~\ref{theorem3}.}
\end{supplement}

% imsref loaded by akundreckaite, 2011-05-04 14:56:29

\printaddresses

\end{document}